\documentclass{article}
\usepackage{graphicx}
\usepackage{framed}
\usepackage[normalem]{ulem}
\usepackage{amsmath}
\usepackage{amsthm}
\usepackage{amssymb}
\usepackage{amsfonts}
\usepackage{enumerate}
\usepackage{graphicx} 
\usepackage{hyperref}
\hypersetup{
  linkcolor  = blue,
  citecolor  = blue,
  urlcolor   = blue,
  colorlinks = true,
} 
\usepackage{natbib}
\usepackage{setspace}
\usepackage{url} 
\usepackage{caption}
\newtheorem{theorem}{Theorem}[section]
\newtheorem{corollary}{Corollary}[theorem]

\newtheorem{proposition}[theorem]{Proposition}

\title{\bf Comparing baseball players across eras via novel Full House Modeling}

\author{Shen Yan$^1$, Adrian Burgos Jr.$^2$, Christopher Kinson$^1$, \\and Daniel J. Eck$^1$\thanks{email addresses: Shen Yan (\url{shenyan3@illinois.edu}); Adrian Burgos Jr. (\url{burgosjr@illinois.edu}); Christopher Kinson (\url{kinson2@illinois.edu}); Daniel J. Eck (\url{dje13@illinois.edu}) }
}

\begin{document}
\maketitle

{\centering
1. Department of Statistics, University of Illinois Urbana-Champaign \\
2. Department of History, University of Illinois Urbana-Champaign \\}

\def\spacingset#1{\renewcommand{\baselinestretch}%
{#1}\small\normalsize} \spacingset{1}


\bigskip
\begin{abstract}
A new methodological framework suitable for era-adjusting baseball statistics is developed in this article. Within this methodological framework specific models are motivated. We call these models Full House Models. Full House Models work by balancing the achievements of Major League Baseball (MLB) players within a given season and the size of the MLB talent pool from which a player came. We demonstrate the utility of Full House Models in an application of comparing baseball players' performance statistics across eras. Our results reveal a new ranking of baseball's greatest players which include several modern players among the top all-time players. Modern players are elevated by Full House Modeling because they come from a larger talent pool. Sensitivity and multiverse analyses which investigate the how results change with changes to modeling inputs including the estimate of the talent pool are presented. 
\end{abstract}

\noindent%
{\it Keywords:}  Order statistics, Nonparametric estimation, Extrapolation techniques, Era-adjustment, Sports statistics.
\vfill

\newpage
\spacingset{1.9} 
\section{Introduction}
\label{introduction}

Debating which baseball players are the best of all-time is a fun exercise among baseball fans. There are many statistics for measuring the greatness of players, and these statistics are often used as justification in these debates. Career or peak offensive or pitching statistics such as home runs, batting average, runs batted in, earned run average, strikeouts and pitcher wins used to dominate discussions of baseball players. Sabermetricians and Statisticians investigated more nuanced measures of a player's abilities, see \citep{thorn1984hidden, mccracken2001pitching, albert2003curveball, lewis2003moneyball, tango2007book, warbr}. These investigations led to the creation of on-base plus slugging percentage, weighted on-base average,  field independent pitching.  Various adjustments to these statistics to account for stadium effects, league effects, and batted ball profile of a batter or batted ball profile allowed of a pitcher have led to even better measures of a player's ability. One statistic in particular that has captured the imagination of baseball fans is wins above replacement, abbreviated as WAR \citep{baumer2014statistician, warbr}. WAR is an attempt by the sabermetric baseball community to summarize a player’s total contributions to their team in one statistic \citep{slowinski2010war}. There are several proprietary versions of WAR that exist with Baseball Reference WAR, often denoted bWAR or rWAR \citep{forman2010bWAR}, and Fangraph's WAR, denoted fWAR \citep{slowinski2010war},  and WARP \citep{bp2013warp} being the most popular. In addition to these versions of WAR, there exists openWAR that is based on public data and careful transparent methodology \citep{BB15}. There is no shortage of statistics for individuals to quote when they passionately argue for who they think are the greatest baseball players ever.

Debates about the all-time greatest baseball players often involve comparisons of players who accrued their statistics in vastly different eras. This makes for lively debate as people present their cases for or against the relative merit of statistics from different time periods. There are many factors that contribute to the differences of eras that are separated in time. Some factors include population size, Westward and Southern expansion of the USA and baseball, league expansion and team relocation, changing interest levels in baseball, erosion of the Minor Leagues, changing approaches to playing baseball, changing managerial strategies, integration, globalization, wars, international relations, changing salaries, changing modes of transportation, changing training methods, medical technology, changing media environments, etc. \citep{land1994organizing, Gould96, schmidt2005concentration, burgos2007playing, rader2008baseball, armour2016baseball, DE20, DE20b}. A starting place for confronting the vast era differences can be found in the work of Stephen Jay Gould. \cite{Gould96} suggested that the entire distribution of achievement, or ``full house of variation," is relevant when considering the achievements of baseball players across eras. This idea was also recognized as the starting place for an answer by \cite{SM13} and \cite{SM16}. These works consider seasonal standard deviations as a proxy for changing eras. \cite{BRL99} assumed that the changing talent pool is captured by seasonal random effects. \cite{BRL99} also considered overlapping player aging curves as a way to establish bridges across eras. In addition to all of the sophisticated statistics approaches, there are several websites and books which provide their own lists of the greatest baseball players using a combination of statistics and opinion \citep{ESPN}, variations of WAR \citep{HOS}, and painstaking narrative curation informed by a plethora of statistics and context considerations \citep{posnanski2021baseball}.

Our contribution to these discussions is a novel modeling approach that extends Gould's ``full house of variation'' concept to the talent pool. This method works by using an estimate of the size of the talent pool as a modeling input and connecting order statistics of observed MLB statistics to individuals in the talent pool. With distributional assumptions placed on the process that assigns latent talent to individuals in the talent pool, an assumption that the most talented people are in the MLB, and a pairing between observed statistics and latent talent such that the best performer according to a particular metric is paired with the highest latent talent, the second best performer is paired with the second most talented individual and so on, we can estimate the latent talent values of MLB players. These mechanics allow for a balancing between how far one stands from their peers and the size of the talent pool. To have a high talent score, one must stand out from their peers in their own time and be a product of a large talent pool. Historically great players like Babe Ruth, Ty Cobb, Lefty Grove, and Walter Johnson, to name a few, dwarfed their contemporaries by such a degree that they remain among the all-time greatest players despite playing in eras that we estimate to have small talent pools relative to more modern eras.

We argue that our results indicate an era-neutral ranking of baseball players. However, it is very important to note that this claim is built on the assumption that we have properly estimated the talent pool. Our estimation of the talent pool is discussed in this article and in our Supplementary Materials. Our rankings reveal that more modern baseball players now sit atop the lists of baseball's greatest than many other approaches. For example, our era-adjusted versions of bWAR (ebWAR) and fWAR (efWAR) favor Barry Bonds and Willie Mays over Babe Ruth. A detailed investigation into how we estimated the talent pool for each season will be provided. Sensitivity analyses, multiverse analyses \citep{steegen2016increasing}, and alternate talent pool estimates will be considered. We now present Full House Modeling through an application of comparing baseball players across eras.

\section{Full House Model Setting} 
\label{setting}

Let $N_i$ denote the size of the talent pool in year $i$. We will suppose that every individual $j = 1,\ldots, N_i$ has an underlying talent value $X_{i,j} \stackrel{\text { iid }}{\sim} F_{X}$. We denote $X_{i,(j)}$ as the $j$th ordered talents in year $i$.
We define $g_{i}\left(\cdot, \cdot\right)$ as the MLB inclusion function, where $g_{i}\left(X_{i,j}, \textbf{X}_{i,-j}\right)=1$ indicates that individual $j$ is an active MLB player in the $i$th year, and $g_{i}\left(X_{i,j}, \textbf{X}_{i,-j}\right)=0$ indicates that individual $j$ is not an active MLB player in the $i$th year. Let $\textbf{X}_{i,-j}$ be the vector of all individual talents not including the component $j$. We will assume that the most talented individuals from the talent pool in year $i$ are active players in the MLB so that 
\begin{equation} \label{eq:MLBselection}
	g_{i}\left(X_{i, j}, \mathbf{X}_{i,-j}\right) = 1\left(X_{i,j} \geq X_{i,(N_i-n_i + 1)}\right),
\end{equation}
where $n_i$ is the number of active MLB players in year $i$ and $1(\cdot)$ is the indicator function. Note that: 1) our estimation of the talent pool accounts for changing interest in baseball over time; 2) it is likely that there are potential players not in the MLB that have more talent in a certain area than some active MLB players. That being said, we think it is sensible to assume that those players would not be among the most talented players in the MLB if they were to be in the MLB.

We now suppose that in year $i$ an active MLB player $j$ has an observed statistic $Y_{i,j}$. For example, $Y_{i,j}$ can be batting average or wins above replacement (WAR) per game. We suppose that $Y_{i,j} \overset{iid}{\sim} F_{Y_i}$ where $F_{Y_i}$ will be continuous. We denote $Y_{i,(j)}$ as the $j$th ordered statistic for players in the MLB during year $i$.
The key to the Full House Model is connecting talent values $X$ with observed statistics $Y$. This is achieved by assuming that outcome data $Y$ arrives from a pairing with $X$, $(Y_{i,(j)}, X_{i,(N_i-n_i+j))})$, $j = 1,\ldots,n_i$. As an example, the highest performer in the MLB in year $i$ as judged by values $Y_{i,j}$ will be assumed to have the highest talent score in year $i$. More detail on this pairing is given in Sections \ref{parametric} and \ref{nonparametric}.

 \subsection{Parametric distributions for baseball statistics} \label{parametric}

We now demonstrate how the Full House Model works in a parametric setting.
Consider the pair $(Y_{i,(j)}, X_{i,(N_i-n_i+j)})$ and suppose that the distribution corresponding to $Y_{i, j}$ from the $i$th system is known to belong to a continuous parametric family indexed by unknown parameter $\theta_{i},$ and let $F_{Y_i}(\cdot \mid \theta_i)$ be a parametric CDF with parameter $\theta_i \in \mathbb{R}^{p_{Y_i}}$. We can estimate $\theta_i$ with $\hat{\theta_i}$ and plug the estimator into the CDF $F_{Y}(\cdot \mid \hat{\theta_i})$.

In order to connect the  $Y_{i,(j)}$ and $X_{i,(N_i-n_i+j)}$ and obtain an estimate of the underlying talents,  we will make use of the following classical order statistics properties, 
$$
\begin{aligned}
F_{Y_i}\left(Y_{i,(j)} \mid \theta_i\right) & \sim U_{i,(j)}, & & F_{Y_i}\left(Y_{i,(j)} \mid \hat{\theta_i}\right) \approx U_{i,(j)}, \\
F_{Y_{i,(j)}}\left(Y_{i,(j)} \mid \theta_i\right) & \sim U_{i,j}, & & F_{Y_{i,(j)}}\left(Y_{i,(j)} \mid \hat{\theta_i}\right) \approx U_{i,j},
\end{aligned}
$$
where $\approx$ means approximately distributed, $\sim$ means distributed as, $U_{i,j} \sim U(0,1)$, and $U_{i,(j)} \sim \operatorname{Beta}(j, n_i+1-j)$ and the quality of the approximation in the right-hand side depends upon the estimator $\hat{\theta}_i$ and the sample size. 

We now connect the order statistics to the underlying talent distribution that comes from a population with $N_i \geq n_i$ observations when $F_{X}$ is known. This connection is established with 
$
F_{X_{i, (N_i-n_i+j)}}^{-1}\left(F_{U_{i,(j)}}\left(F_{Y_{i}}\left(Y_{i,(j)} \mid \theta_i\right)\right)\right) \sim F_{X_{i, (N_i-n_i+j)}}^{-1}\left(F_{U_{i,(j)}}\left(U_{i,(j)}\right)\right) = X_{i, (N_i-n_i+j)}.
$
This is estimated with
\begin{equation} \label{eq:parest}
	F_{X_{i, (N_i - n_i + j)}}^{-1}\left(F_{U_{i,(j)}}\left(F_{Y_{i}}\left(Y_{i,(j)} \mid \hat{\theta}_i\right)\right)\right) \approx F_{X_{i, (N_i - n_i + j)}}^{-1}\left(F_{U_{i,(j)}}\left(U_{i,(j)}\right)\right) = X_{i, (N_i - n_i + j)}.
\end{equation}

\subsection{Nonparametric distribution for baseball statistics} \label{nonparametric}

\subsubsection{Past methods and challenges of nonparametric approach} \label{challenges}

The empirical cumulative distribution function (CDF) $\widehat{F}_{Y_i}$ is a widely used nonparametric approach in estimating the CDF. However, $\widehat{F}_{Y_i}$ fails in our setting because $\widehat{F}_{Y_i}(Y_{i,(n_i)}) = 1$ which, when mapped to $X$ values through a similar approach to \eqref{eq:parest}, yields $X_{i, (N_i)} = \sup_x\{x : F_X(x) < 1\}$. The implication here is that the highest achiever in year $i$ is estimated to have maximal possible talent, and this is nonsense. Therefore, we have an extrapolation problem. There are many alternatives to $\widehat{F}_{Y_i}$. For example, one could use piecewise linear function estimation \citep{LD98, KL12}, kernel estimation \citep{SB86}, and semi-parametric conjugated estimation \citep{SF95}. One could also consider parametric families of the generalized Pareto distribution that have flexible behavior in both tails \citep{ST20}.


All methods above, and others not mentioned, fit within a more general Full House Modeling paradigm than what we motivate here. However, in the application to baseball data, the range of the distribution is naturally constrained, and outlying achievements are lauded for their rarity. In preliminary analyses we found that several of the above methods led to era-adjusted results that rewarded performances of top achievers that did not stand far above their peers. The main issue for these methods was that a tight grouping of top achievers would result in the very highest achiever having an extremely large estimated talent score when that player only stood a relatively short distance higher than the other top achievers. 
We found that an approach motivated from \cite{SF95} performed well with these issues. Details will be discussed in Section~\ref{Sec:ystar}.

\subsubsection{Handling extrapolation} 
\label{extrapolated}

In the nonparametric setting, we motivate a variant of a natural interpolated empirical CDF as an estimator of the system components distribution $F_{Y_i}$ to solve the problems mentioned in the previous section. We consider surrogate sample points to construct an interpolated version of the empirical CDF $\widetilde{F}_{Y_i}$ and this type of interpolated CDF is a standard technique to replace the empirical CDF \citep{KL12}.

 We construct the interpolated CDF in the following manner: We first construct surrogate sample points 
$\widetilde{Y}_{i,(1)} =Y_{i,(1)}-Y^{*}_i$, 
$\widetilde{Y}_{i,(j)} =\left(Y_{i,(j)}+Y_{i,(j-1)}\right) / 2, j=2, \ldots, n$, and
$\widetilde{Y}_{i,(n_i+1)} =Y_{i,(n_i)}+Y^{**}_i$,
where $Y^*_i$ is the value to construct the lower bound and $Y^{**}_i$ is the value to construct the upper bound. With this construction, we build $\widetilde{F}_{Y}$ as

\begin{equation} \label{eq:empcdf}
\widetilde{F}_{Y}(t) = 
  \sum_{j=1}^{n_i}\left(\frac{j-1}{n_i} + \frac{t-\widetilde{Y}_{i,(j)}}{n_i\left(\widetilde{Y}_{i,(j+1)}-\widetilde{Y}_{i,(j)}\right)}\right) 1\left(\widetilde{Y}_{i,(j)} \leq t<\widetilde{Y}_{i,(j+1)}\right)+1\left(t \geq \widetilde{Y}_{i,(n_i+1)}\right)
\end{equation}

The estimator $\widetilde{F}_{Y}$ is desirable for three reasons. First, we found \eqref{eq:empcdf} to be quick computationally. Second, we do not assume that the observed minimum and observed maximum constitute the actual boundaries of the support of $Y$. Third, $\widetilde{F}_{Y}\left(Y_{i,(1)}\right)$ and $\widetilde{F}_{Y}\left(Y_{i,(n_i)}\right)$ provide reasonable estimates for the cumulative probability at $Y_{i,(1)}$ and $Y_{i,(n_i)}$ by considering their respective value of $Y^*_i$ and $Y^{**}_i$. Here $Y^{**}_i$ is chosen to measure how far the highest achiever in year $i$ stood from their peers where small values of $Y^{**}_i - Y_{i,(n_i)}$ have the interpretation that $Y_{i,(n_i)}$ is an outlying performance and large values of $Y^{**}_i - Y_{i,(n_i)}$ indicate the opposite. 

The approximations to facilitate our methodology are similar to the ones in the parametric case. Latent talent values can be found as follows, 
$
F_{X_{i, (N_i-n_i+j)}}^{-1}\left(F_{U_{i,(j)}}\left(F_{Y_{i}}\left(y_{i,(j)} \right)\right)\right) \sim F_{X_{i, (N_i-n_i+j)}}^{-1}\left(F_{U_{i,(j)}}\left(U_{i,(j)}\right)\right) = X_{i, (N_i-n_i+j)}.
$
This can be estimated as
\begin{equation} \label{eq:nonparest}
	F_{X_{i, (N_i - n_i + j)}}^{-1}\left(F_{U_{i,(j)}}\left(\widetilde{F}_{Y_{i}}\left(y_{i,(j)} \right)\right)\right) \approx F_{X_{i, (N_i - n_i + j)}}^{-1}\left(F_{U_{i,(j)}}\left(U_{i,(j)}\right)\right) = X_{i, (N_i - n_i + j)}.
\end{equation}
Notice that $\widetilde{F}_{Y_i}(t)$ was explicitly constructed to be close to $\widehat{F}_{Y_i}(t)$. We formalize this statement in the Supplementary Materials.

\subsubsection{Choosing $Y^{*}_i$ and $Y^{**}_i$ } 
\label{Sec:ystar}

The cumulative probabilities $\widetilde{F}_{Y}\left(Y_{i,(1)}\right)$ and $\widetilde{F}_{Y}\left(Y_{i,(n_i)}\right)$ are, respectively, functions of $Y^{*}_i$ and $Y^{**}_i$. In this section, we describe the role of $Y^{*}_i$ and $Y^{**}_i$ and how these quantities are chosen in our application to historical baseball rankings. The lower bound $Y^{*}_i$ determines the lower tail behavior of talent distribution, and in fact, most normal or low talents would concentrate in a similar scale or size \citep{NM05}. Therefore $Y^{*}_i$ can be a small positive value, we define $Y^{*}_i = Y_{i, (2)} - Y_{i, (1)}$. This is not the only choice of $Y^{*}_i$, and other reasonable choices can also be set to the $Y^{*}_i$.
 
We now discuss how we calculate $Y^{**}_i$. Our approach follows a simple nonparametric tail extrapolation method motivated in \cite{SF95}. First, some preliminaries. 
The $p$-quantile $y_{p}$ of $F$ is defined as the smallest value for which $F\left(y_{p}\right)=p$, i.e. $y_{p}=\inf\{y: F(y) \geq p\}$. Hence $P\left(Y_{i,j} \leq y_{p}\right)=p$. Now suppose that $p_{i,j,\gamma, n_i}$ is the value of $p$ that satisfies $\gamma=P\left(Y_{i,(j)} \geq y_{p}\right)$. For each $j$ and $\gamma$ there is an approximation of $p_{i,j,\gamma, n_i}$ for choices of $\gamma$ \citep{SF95}. \cite{HD83} gave a specific justified approximation of $p_{i,j,\gamma, n_i}$ when $\gamma = 0.5$. This approximation is,
$$
  p_{i,j, .5, n_i} \approx \frac{j-\frac{1}{3}}{n_i+\frac{1}{3}}.
$$ 
With this approximation, we have a tractable means to connect the order statistics $Y_{i,(j)}$ to percentile values $p_{i,j, .5, n_i}$ corresponding to the median value of a quantile $y_{p_{i,j, .5, n_i}}$.

Following \cite{SF95}, we consider the regression fit on points $(h(p_{i,j,.5,n_i}), Y_{i,(j)})$, $j = n_i - k + 1,\ldots,n_i$, where $k$ is the number of extreme data values to use in the extrapolation step, and $h$ is a function of tail probabilities. The specific choices that we considered for $h$ to model tail probabilities are similar to those in Section 2 of \cite{SF95}: 
\begin{itemize}
\item linear transformation: $h_{\theta}(p) = \theta_1 + \theta_2 p$;
\item logit transformation: $h_{\theta}(p)  = \theta_1 + \theta_2 \log(p/(1-p))$;
\end{itemize}
In practice, we chose $h$ among the candidates by selecting whichever $h$ maximized regression fit as judged by $R^2$ values. 

The value of $k$ is determined by a similar procedure to that in Section 5 of \cite{SF95}: we specify that $k \in [K_1,K_2]$ where $K_{1}=\max (6,\lfloor 1.3 \sqrt{n}\rfloor)$ and $K_{2}=2\left\lfloor\log_{10}(n) \sqrt{n}\right\rfloor$. We then choose $k$ as the value that maximizes adjusted $R^2$ values among regressions motivated in Sections 3-4 of \cite{SF95}. We found that the regression approach in Section 3-4 of \cite{SF95} suffered similar problems as those mentioned in the the last paragraph of Section~\ref{challenges}.

With $k$ and $h$ selected, we then compute $Y_i^{**}$ through a connection between $\widetilde{F}_{Y_i}(Y_{i,(n_i)})$ and the regression fit on points $(h(p_{i,j,.5,n_i}), Y_{i,(j)})$, $j = n_i -k + 1,\ldots,n_i$. Specifically, we find $Y_i^{**}$ as the solution of the following optimization problem 
\begin{equation} \label{eq:optim}
  Y_i^{**} = \text{argmin}_y\Big|h^{-1}(Y_{i,(n_i)}) - \widetilde{F}_{Y_i}(Y_{i,(n_i)};y)\Big|,	
\end{equation}
where $\widetilde{F}_{Y_i}(\cdot;y)$ is $\widetilde{F}_{Y_i}$ defined with respect to a value $y$ replacing $Y_i^{**}$ in its construction. The intuition of \eqref{eq:optim} for large outlying values of $Y_{i,(n_i)}$ is as follows: the value of $h^{-1}(Y_{i,(n_i)})$ corresponds to percentile $p_{i,n_i,.5,n_i}$ that is close to $1$, and this pulls $Y_i^{**}$ towards zero so that $\widetilde{F}_{Y_i}(Y_{i,(n_i)})$ is also close to $1$.

\subsection{Estimate how players will perform in a different year} \label{estimation}

We can now reverse engineer the process above to obtain era-adjusted statistics in any context that is desired.
Consider the the pair $(Y_{i,(j)}, X_{i,(N_i-n_i+j)})$ in the $i$th year. We first put the talent value $X_{i,(N_i-n_i+1)}$ obtained by \eqref{eq:parest} or \eqref{eq:nonparest} in the new talent pool of desired year $m$ and reverse the process to obtain the hypothetical baseball statistics in year $m$, which we will denote as $Y_{i,j,m}$. The distribution $F_X$ is known. 

More formally, $Y_{i,j,m}$ are computed as follows when baseball statistics are estimated parametrically:
\begin{equation} \label{eq:parrev}
	Y_{i,j,m} = F_{Y_{m}}^{-1} \left( F^{-1}_{U_{m, (l_{i,j,m})}}
\left(F_{X_{m, (N_m - n_m + l_{i,j,m}))}}\left(X_{i, (N_i-n_i+j)}  \right) \right)| \hat{\theta}_m\right),
\end{equation}
where $l_{i,j,m}$ is the rank of $X_{i,(N_i-n_i+j)}$ among the values \\
$
\left\{X_{i,(N_i-n_i+j)}, X_{m,\left(N_m-n_m+t)\right)}: t=1, \ldots, n_{m}\right\}.
$
The baseball statistics $Y_{i,j,m}$ are computed as follows when baseball statistics are estimated nonparametrically:
\begin{equation} \label{eq:nonparrev}
	Y_{i,j,m} = \widetilde{F}_{Y_{m}}^{-1} \left( F^{-1}_{U_{m, (l_{i,j,m})}}
\left(F_{X_{m, (N_m - n_m + l_{i,j,m})}}\left(X_{i, (N_i - n_i + j)} \right) \right)\right).
\end{equation}

\subsection{Putting it all together}
\label{Sec:algorithm}

We conclude Section \ref{setting} by expressing the working mechanics of the Full House Model in an algorithmic format. Steps 1-3 describe how one obtains talents $X$ from observations $Y$. Steps 4-6 describe how one reverse-engineers the process to obtain new $Y$ values in a new context from the talents in $X$ computed in step 3. This algorithm is presented below: 

\begin{itemize}
	\item[Step 1:] Input the statistics $Y_{i,1}, Y_{i,2}, \dots, Y_{i,n_i}$, the talent pool size $N_i$, and the number of active MLB players $n_i$ for year $i$. Declare latent distribution $X \sim F_X$ and system inclusion mechanism $g$ \eqref{eq:MLBselection}.
	\item[Step 2:] Sort the components yielding $Y_{i,(1)}, Y_{i,(2)}, \dots, Y_{i,(n_i)}$.
	\item[Step 3.] Obtain talent scores parametrically \eqref{eq:parest} or nonparametrically \eqref{eq:nonparest}.
	\item[Step 4.] Declare a setting for which hypothetical statistics $Y_{i,j,m}$ are desired. 
	\item[Step 5.] Apply steps 1-3 to extract talent scores for active MLB players in year $m$, $\{X_{m,\left(N_m-n_m+t)\right)}: t=1, \ldots, n_{m}\}$, and find the rank $l_{i,j,m}$ of $X_{i,(N_i-n_i+j)}$ among these talent scores.
	\item[Step 6.] Obtain statistics $Y_{i,j,m}$ parametrically \eqref{eq:parrev} or nonparametrically \eqref{eq:nonparrev}.	
\end{itemize}

\begin{figure}[htb!]
\begin{center}
\includegraphics[scale = 0.38]{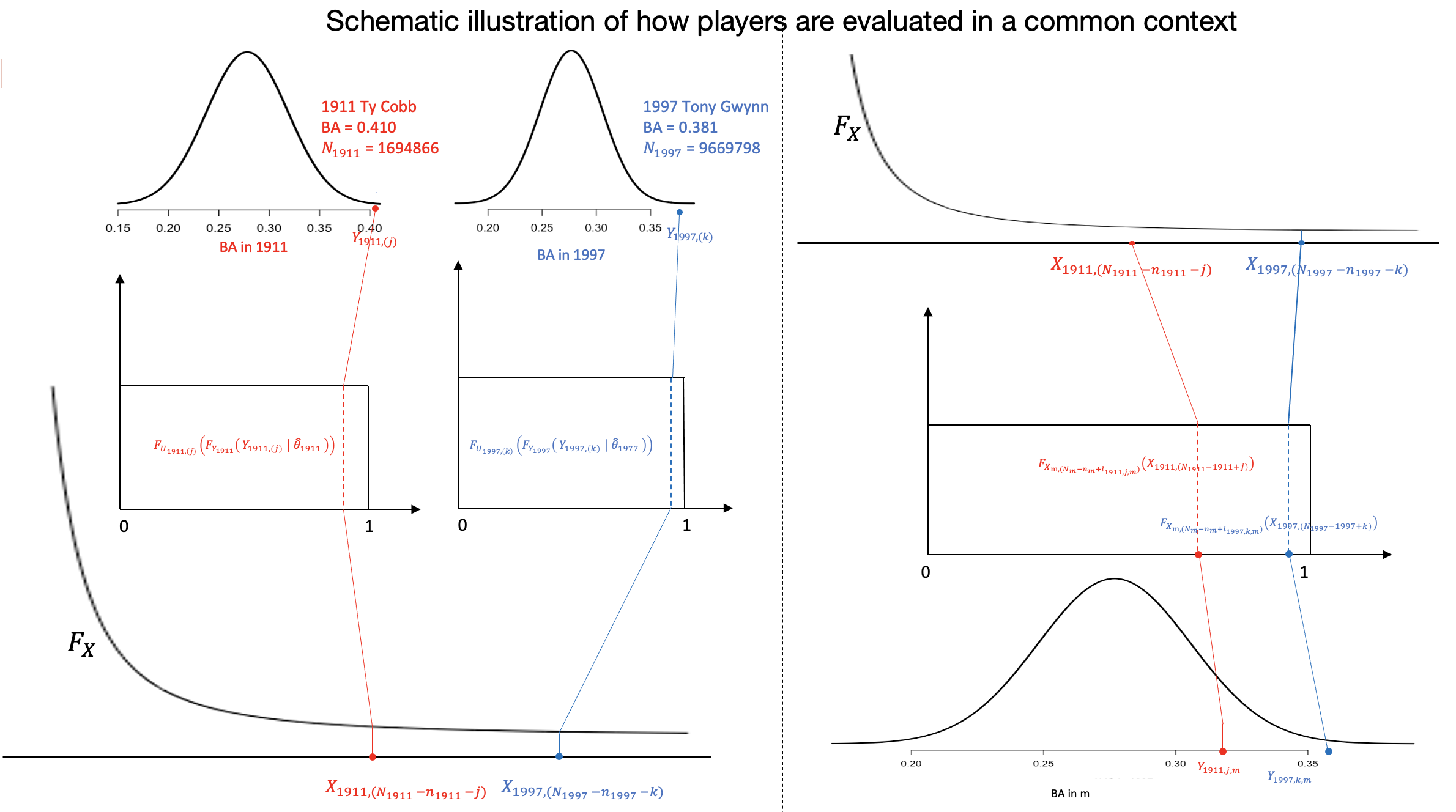}
\end{center}	
\caption{Obtaining era and park-factor adjusted batting averages for two players in different seasons via the Full House Model. The left panel illustrates how to extract the BA talent scores for Ty Cobb in 1911 and Tony Gwynn in 1997 using steps 1-3 in the Algorithm in Section~\ref{Sec:algorithm}. The right panel shows how to compute the era and park-factor adjusted BA in a hypothetical season $m$ for these players using steps 4-6 in the Algorithm in Section~\ref{Sec:algorithm}. Park-factor-adjusted batting averages are assumed to be normally distributed. }
\label{Fig:Illustration}
\end{figure}

The steps of the above algorithm are presented in Figure \ref{Fig:Illustration}. This figure depicts the process of obtaining era and park-factor-adjusted batting average (BA) for two league-leading players, Ty Cobb and Tony Gwynn. We can see even though Ty Cobb's BA in 1911 is larger than Tony Gwynn's BA in 1997, Tony Gwynn's BA talent score is higher. In our Supplementary Materials we conducted a multiverse analysis \citep{steegen2016increasing} which compared Ty Cobb's BA in 1911 and Tony Gwynn's 1997 BA under multiple configurations of our modeling assumptions and data pre-processing regimes. In this analysis, it was found that the vast majority of configurations favored Tony Gwynn over Ty Cobb, and configurations that led to the opposite conclusion involved modeling assumptions that are not satisfied.

\section{Modeling assumptions and data considerations}
\label{Sec:modeling}

\subsection{Historical talent pool}
\label{sec:MLBpop}

An estimate of the talent pool is a central input needed for our model, and our results follow from it. Here we detail the general approach that we employed to estimate the talent pool. Full details can be found here: 
\begin{center}
\url{https://htmlpreview.github.io/?https://github.com/ecklab/era-adjustment-app-supplement/blob/main/writeups/MLBeligiblepop.html}
\end{center}

Our estimate of the talent pool will be pegged to the population of aged 20-29 males from the North East (NE) and Midwest (MW) region of the United States of America. Simply stated, the talent pool will be 
$$
  \frac{\text{NE and MW region population of age 20-29 males}}{\text{proportion of MLB players from NE and MW regions}} \times \text{adjustment}.
$$
Baseball started in the NE and MW regions. Our estimation approach tracks the talent pool as baseball expands throughout the USA and abroad. We adjust for the effects of changing interest in baseball, wars, segregation, and gradual integration of the MLB. 

The proportion of MLB players from the NE and MW regions can be calculated from using birthplace information from the Lahman R package \citep{Lahman}. We estimate the talent pool from the NE and MW regions using US Census data and linear interpolation for years in which we could not find data. We have adjusted population sizes to account for the effects of wars, the rates of integration in the two traditional leagues which comprise the MLB \citep{armour2007effects}, and the changing interest in baseball as measured by Gallup and Harris (links to specific polls are provided in the link at the beginning of this section). Baseball interest is lagged by 10 years to reflect the level of baseball interest that was present when those in the MLB were coming of age. 

\begin{figure}
\includegraphics[scale=0.5]{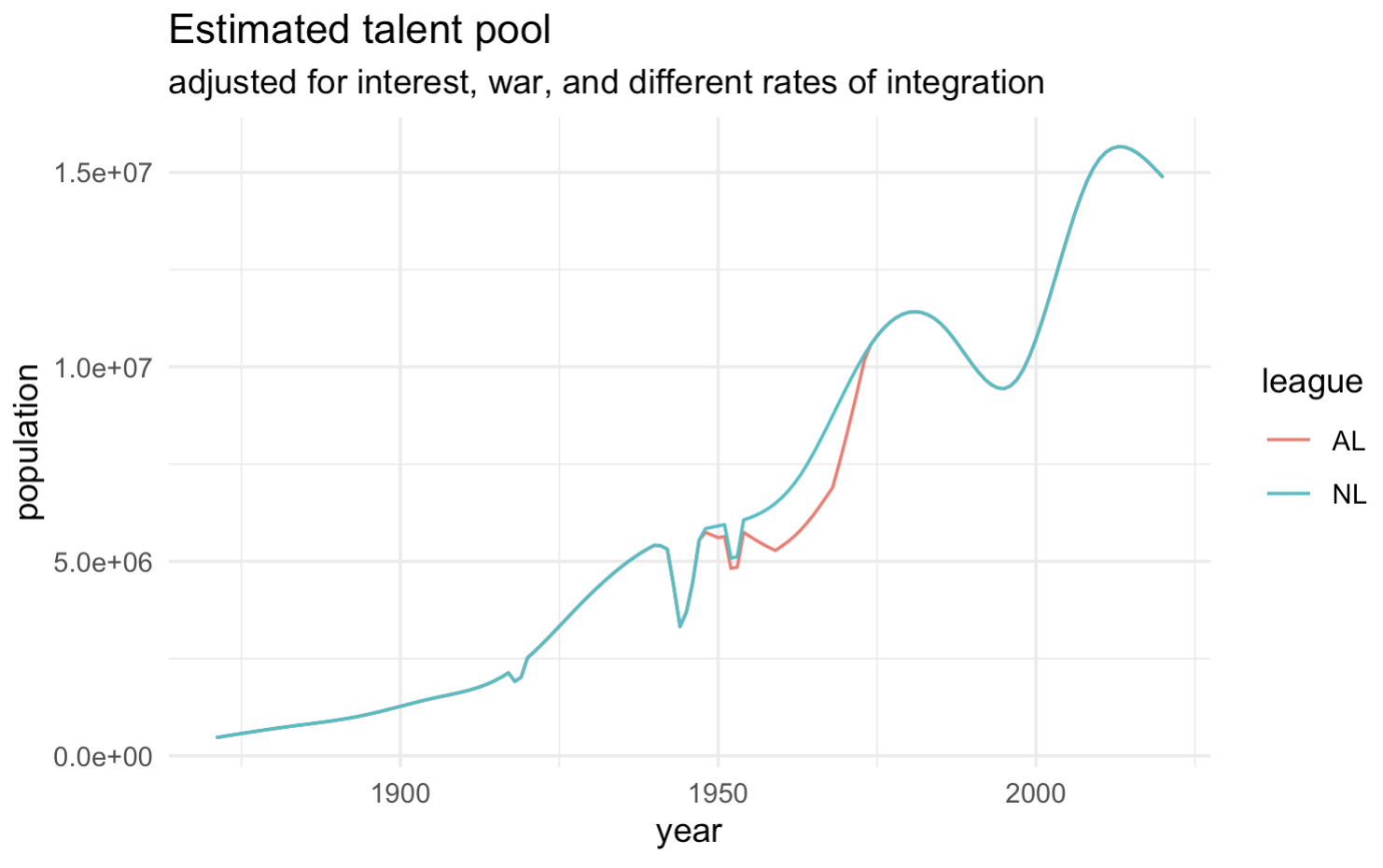}
\caption{Estimated baseball talent pool over time. }
\label{Fig:talentpool}	
\end{figure}

\begin{figure}
\includegraphics[scale=0.5]{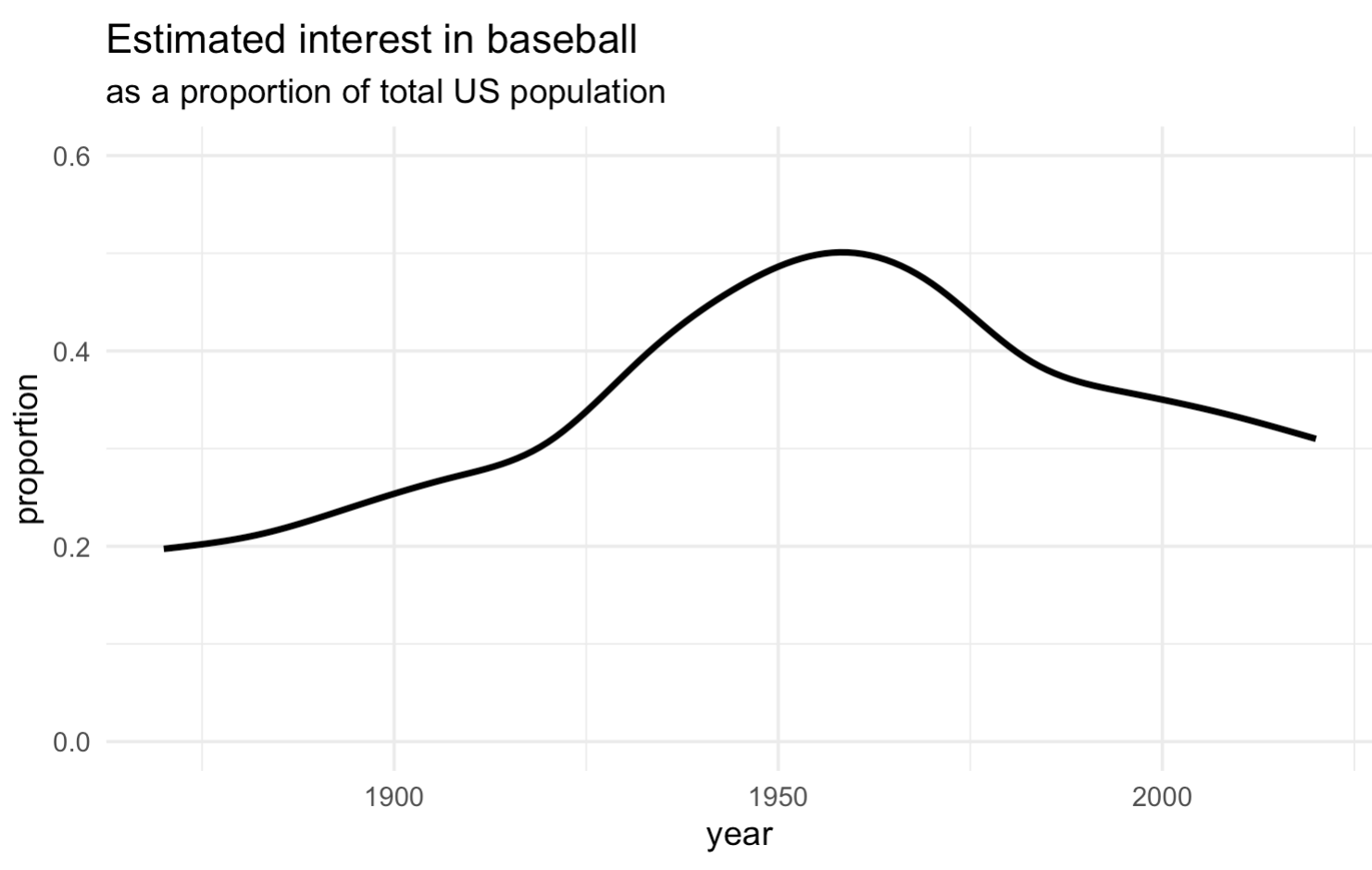}
\caption{Estimated interest in baseball over time. }
\label{Fig:interest}	
\end{figure}

We define baseball interest to be the average of the proportion of people who are generally interested in baseball and the proportion of people who list baseball as their favorite sport. Including general interest in baseball is sensible because not everyone who plays baseball lists baseball as their favorite sport, for example, Tony Gwynn loved basketball before baseball\footnote{\url{https://www.mlb.com/news/tony-gwynn-drafted-in-baseball-basketball-on-same-day}}. Averaging these two sources of baseball interest yields an estimate of the talent pool that aligns with a similarly constructed talent pool formed from select Latin American countries that are smaller than the US in population but are more interested in baseball. Figures~\ref{Fig:talentpool} and \ref{Fig:interest}, respectively, display our estimate of the talent pool and baseball interests over time. 

\begin{figure}
\includegraphics[scale=0.52]{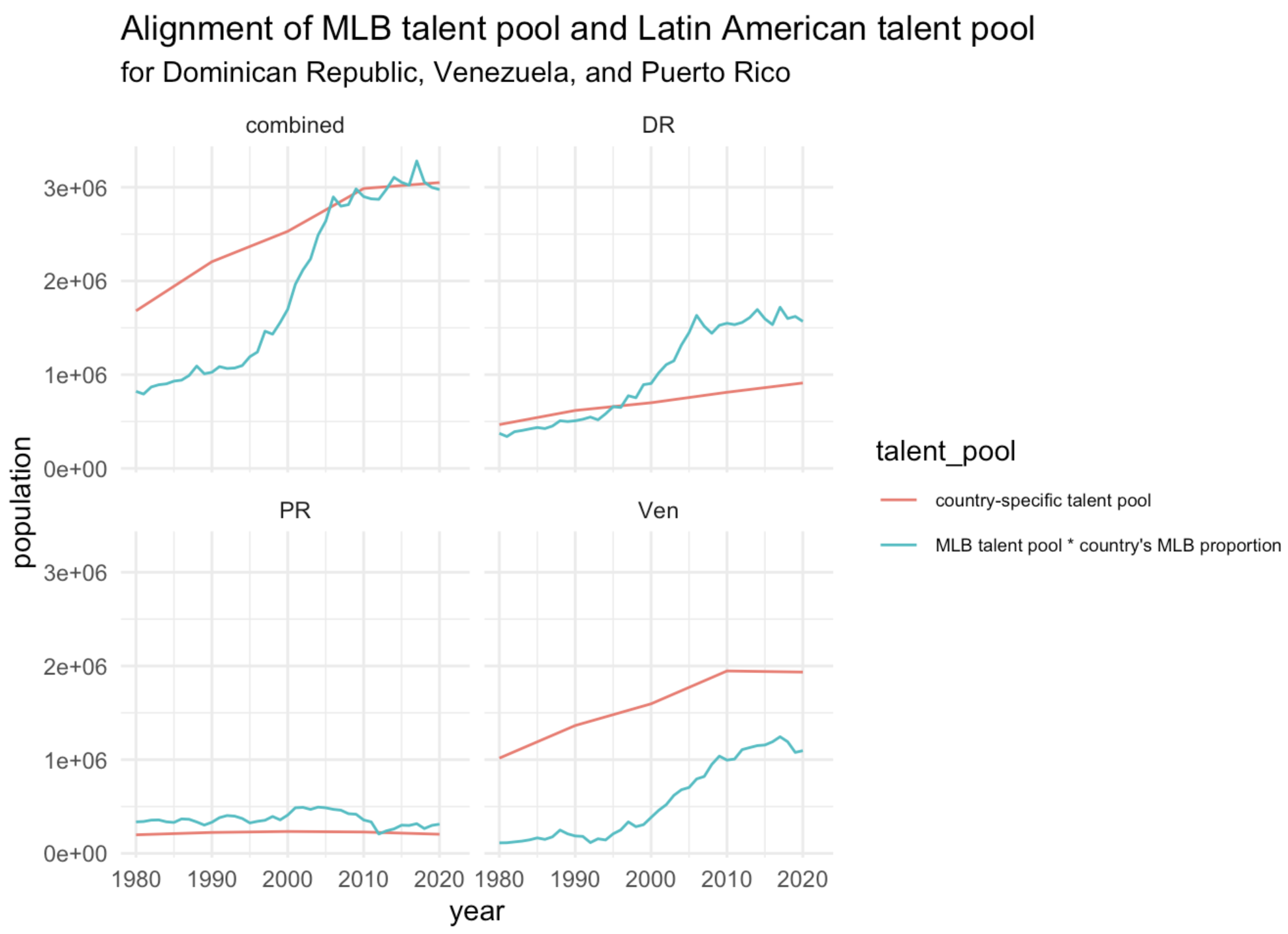}
\caption{MLB representation vs talent pool size for select Latin American countries. The countries considered are the Dominican Republic (DR), Puerto Rico (PR), and Venezuela (Ven), and the combination of these three countries. The blue line (MLB representation) is our estimated talent pool multiplied by the proportion of MLB players from these Latin American countries. The red line (eligible pop) is the estimated talent pool for these Latin American countries. }
\label{Fig:MLBLA}	
\end{figure}

Figure~\ref{Fig:MLBLA} shows that our estimate of the talent pool aligns with a similar talent pool computed from the combination of the Dominican Republic, Puerto Rico, and Venezuela which are the three countries with the most historical MLB players after the United States. This alignment occurs from around 2005 to the present. Around 2005, a large influx of Latin American players began to level off \citep{armour2016baseball}. Separately, the Dominican Republic and Puerto Rico exhibit over-representation in the MLB while Venezuela exhibits under-representation (see Figure~\ref{Fig:MLBLA}). Both of these findings are pronounced but are hardly surprising considering the extensive development of baseball academies in the Dominican Republic and Puerto Rico, and the recent relations between the USA and Venezuela.

\subsection{Modeling assumptions and specifications}

In Section~\ref{setting} we supposed that the all $N_i$ individuals in population $i$ have talents $X_{i,1},\ldots, X_{i,N_i} \overset{iid}{\sim} F_X$. Here $N_i$ corresponds to the size of the talent pool in year $i$. The Full House Model results displayed throughout Section~\ref{rankings} assumed that $F_X$ corresponds to a Pareto($\alpha$) distribution with $\alpha = 1.16$. This choice of $\alpha$ reflects the Pareto principle or ``80-20 rule" \citep{PV64}. \cite{BD10} noted that the superstars are really important for their teams to win and stated that about 80\% of wins appear to be produced by the top 20\% of players. Thus there is some precedent for our assumption on $F_X$. In the Supplementary Materials, we perform a sensitivity analysis on choices of $F_X$. This sensitivity analysis reveals that our conclusions do not depend on the distribution assumed for $F_X$, as expected since the ranking of players is preserved. All baseball statistics will be analyzed using our nonparametric method articulated throughout Section~\ref{nonparametric}. Handedness-specific park-factor adjustments are made to batting averages and home run rates. These adjustments are computed using the same methods in \cite{SM16}.

The Full House Model results displayed throughout Section~\ref{rankings} were estimated in a common context. Namely, we built a common mapping environment to evaluate all player seasons. \cite{SM16} inspired us to use the National League (NL) seasons from 1977 through 1989, excluding the 1981 strike-shortened season, as the seasons comprising the common mapping environment. 
\cite{SM16} noted that the 1977-1989 NL represents possibly the most steady time in baseball history for all of the fundamental offensive occurrences. Additionally, no expansion occurred over these years.
We fit an isotonic regression model with observed performance statistics on estimated talent scores obtained from our Full House Model for all 1977-1989 NL player seasons, excluding the 1981 strike-shortened season. We obtained predicted performance statistics (which we will denote as $\hat{Y})$ from this isotonic regression model to pair with estimated talent scores. 

We set the number of players in the common mapping environment to be equal to the maximum number of full-time players (denoted $n$) throughout the 1871-2023 seasons. Then we select as data pairs to form the common mapping environment as $(\hat{Y}_{(k)}, X_{(k)})$, $k = 1,\ldots, n$ corresponds to the $\frac{k}{n}$th quantile of all $(\hat{Y},X)$ data pairs that formed the common mapping environment. 
From here, steps 1-3 of the algorithm in Section~\ref{Sec:algorithm} were applied to all MLB player seasons individually, and steps 4-6 of the algorithm in Section~\ref{Sec:algorithm} found era-adjusted statistics within the context of the common mapping environment. 

Further details about pre-processing and adjustments to batting statistics and pitching statistics can be found in the Supplementary Materials.

\section{Results} 
\label{rankings}

We present our era-adjusted rankings of baseball players. Tables~\ref{Tab:ecareerbatters} and \ref{Tab:ecareerpitchers} display the top 25 career era-adjusted rankings for, respectively, batters and pitchers across several key statistics. Table~\ref{Tab:ecareercombined} displays top-25 combined era-adjusted WAR and JAWS lists for batters and pitchers. JAWS is the average of a players' career WAR and the players' total WAR from his seven best seasons. Note that Babe Ruth's WAR and JAWS combine his batting and pitching contributions. Table~\ref{Tab:epeakbatting} displays top-25 era-adjusted four-year peak batting averages and at-bats per home run rates.

Our rankings favor more modern players who come from a larger talent pool than legendary players of the past. We find that Barry Bonds and Willie Mays move to the top of WAR-based rankings lists for batters, and Roger Clemens, Greg Maddux, and Randy Johnson move to the top of WAR-based ranking lists for pitchers. Interestingly, Lefty Grove, who dominated during an era that was favorable to batters, jumps a few spots on WAR-based ranking lists. Moreover, Lefty Grove paces those above him on a rate-basis (ebWAR or efWAR per innings pitched). Several active players at the time of this writing populate our era-adjusted top-25 ranking lists. And the legendary players of the past do not disappear. For example, Babe Ruth is the career leader in home runs, has the third highest at bats per home run four-year peak, and is fourth in career ebWAR, efWAR, ebJAWS, and efJAWS among all players. 

Our career and peak era-adjusted home run ranking lists reward modern players because they are products of a larger relative talent pool than past players, but it penalizes players from the steroids era because players' accomplishments are judged relative to their peers. Our career and peak era-adjusted batting ranking lists reward top players from the era of the pitcher (1962-1968) who achieved relatively pedestrian batting averages and home runs due to conditions that favored pitchers such as expanded strike zones and high pitching mounds. Our peak batting average ranking rewards present-day players (at the time of this writing) who stand far from their peers at a time in which batting averages are much lower than in past eras (for example, Luis Arreaz's 2023 BA of 0.354 is the only season in which a qualified batter eclipsed 0.350 BA since Josh Hamilton had a 0.359 BA in 2010). In our Supplementary Materials we report top-25 career and four-year peak batting average ranking list resulting from the assumption that batting averages for full-time players follows a normal distribution. These lists are broadly similar to what is seen in Table~\ref{Tab:epeakbatting} with notable elevation to older era-players and declines to present-day players (at the time of this writing). Also included in our Supplementary Materials is an analysis which demonstrates that a normal distribution assumption placed on batting averages for full-time players does not hold up to Shapiro-Wilk testing \citep{SS65} with the tests overly rejecting nominal tolerance. For these reasons we have reported batting average results that arise when no distribution is specified for batting averages.

\begin{table}[ht]
\scriptsize
\centering
\begin{tabular}{rlrlrlrlr}
  \hline
 & name & ebWAR & name & efWAR & name & HR & name & BA \\ 
  \hline
1 & Barry Bonds & 153.89 & Barry Bonds & 145.24 & Babe Ruth & 702 & Tony Gwynn & 0.342 \\ 
  2 & Willie Mays & 144.08 & Willie Mays & 135.39 & Henry Aaron & 689 & Rod Carew & 0.329 \\ 
  3 & Henry Aaron & 135.60 & Henry Aaron & 128.05 & Albert Pujols & 662 & Ichiro Suzuki & 0.327 \\ 
  4 & Babe Ruth & 127.29 & Babe Ruth & 120.44 & Barry Bonds & 654 & Jose Altuve & 0.327 \\ 
  5 & Alex Rodriguez & 120.29 & Stan Musial & 113.03 & Reggie Jackson & 578 & Ty Cobb & 0.320 \\ 
  6 & Stan Musial & 119.51 & Alex Rodriguez & 110.30 & Willie Mays & 577 & Roberto Clemente & 0.320 \\ 
  7 & Ty Cobb & 114.48 & Ty Cobb & 108.77 & Mike Schmidt & 561 & Miguel Cabrera & 0.320 \\ 
  8 & Albert Pujols & 111.86 & Ted Williams & 107.75 & Alex Rodriguez & 547 & Joe DiMaggio & 0.318 \\ 
  9 & Mike Schmidt & 109.58 & Mike Schmidt & 106.41 & Frank Robinson & 535 & Wade Boggs & 0.316 \\ 
  10 & Rickey Henderson & 109.08 & Rickey Henderson & 103.90 & Ken Griffey Jr & 528 & Buster Posey & 0.316 \\ 
  11 & Ted Williams & 107.86 & Albert Pujols & 97.34 & Willie Stargell & 528 & Mike Trout & 0.315 \\ 
  12 & Tris Speaker & 102.26 & Joe Morgan & 96.07 & David Ortiz & 521 & Ted Williams & 0.314 \\ 
  13 & Joe Morgan & 100.17 & Frank Robinson & 95.92 & Willie McCovey & 515 & Freddie Freeman & 0.314 \\ 
  14 & Frank Robinson & 99.93 & Mel Ott & 95.72 & Harmon Killebrew & 508 & Joe Mauer & 0.314 \\ 
  15 & Mel Ott & 99.74 & Tris Speaker & 95.13 & Ted Williams & 503 & Stan Musial & 0.313 \\ 
  16 & Cal Ripken Jr & 97.39 & Rogers Hornsby & 94.42 & Mickey Mantle & 502 & Willie Mays & 0.312 \\ 
  17 & Rogers Hornsby & 97.01 & Mickey Mantle & 94.30 & Eddie Mathews & 502 & Bill Terry & 0.312 \\ 
  18 & Lou Gehrig & 95.87 & Cal Ripken Jr & 93.24 & Eddie Murray & 498 & Robinson Cano & 0.311 \\ 
  19 & Mickey Mantle & 95.37 & Lou Gehrig & 92.98 & Jimmie Foxx & 493 & Henry Aaron & 0.310 \\ 
  20 & Carl Yastrzemski & 95.20 & Carl Yastrzemski & 92.64 & Stan Musial & 492 & Derek Jeter & 0.310 \\ 
  21 & Adrian Beltre & 95.01 & Honus Wagner & 89.81 & Dave Winfield & 491 & Vladimir Guerrero & 0.310 \\ 
  22 & Wade Boggs & 92.51 & Wade Boggs & 87.91 & Mark McGwire & 489 & Al Oliver & 0.310 \\ 
  23 & Roberto Clemente & 91.37 & Mike Trout & 87.87 & Jim Thome & 484 & Matty Alou & 0.310 \\ 
  24 & Eddie Collins & 90.94 & Eddie Mathews & 86.38 & Miguel Cabrera & 480 & Lou Gehrig & 0.309 \\ 
  25 & Mike Trout & 90.43 & Adrian Beltre & 86.34 & Lou Gehrig & 479 & Edgar Martinez & 0.309 \\ 
   \hline
\end{tabular}
\caption{Top 25 batting careers according to era-adjusted bWAR (ebWAR), era-adjusted fWAR (efWAR), era-adjusted home runs, and era-adjusted batting average (minimum 5000 adjusted at-bats).}
\label{Tab:ecareerbatters}
\end{table}

\begin{table}[ht]
\scriptsize
\centering
\begin{tabular}{rlrlrlrlr}
  \hline
 & name & ebWAR & name & efWAR & name & ERA & name & K \\ 
  \hline
1 & Roger Clemens & 145.88 & Roger Clemens & 141.25 & Clayton Kershaw & 2.43 & Nolan Ryan & 6026 \\ 
  2 & Greg Maddux & 113.66 & Greg Maddux & 120.73 & Pedro Martinez & 2.61 & Randy Johnson & 5136 \\ 
  3 & Randy Johnson & 110.81 & Randy Johnson & 109.77 & Greg Maddux & 2.77 & Roger Clemens & 4752 \\ 
  4 & Tom Seaver & 104.31 & Nolan Ryan & 108.30 & Lefty Grove & 2.78 & Steve Carlton & 4221 \\ 
  5 & Lefty Grove & 102.54 & Bert Blyleven & 101.82 & Roger Clemens & 2.81 & Walter Johnson & 3888 \\ 
  6 & Justin Verlander & 100.23 & Steve Carlton & 100.34 & Justin Verlander & 2.81 & Bert Blyleven & 3785 \\ 
  7 & Bert Blyleven & 97.69 & Lefty Grove & 98.80 & Max Scherzer & 2.83 & Tom Seaver & 3656 \\ 
  8 & Phil Niekro & 94.37 & Justin Verlander & 95.07 & Roy Halladay & 2.85 & Don Sutton & 3575 \\ 
  9 & Clayton Kershaw & 93.78 & Gaylord Perry & 94.45 & Randy Johnson & 2.90 & Max Scherzer & 3506 \\ 
  10 & Walter Johnson & 91.53 & Walter Johnson & 91.80 & Tom Seaver & 2.91 & Greg Maddux & 3473 \\ 
  11 & Warren Spahn & 91.20 & Cy Young & 91.28 & Cole Hamels & 2.94 & Gaylord Perry & 3366 \\ 
  12 & Max Scherzer & 90.63 & Tom Seaver & 90.78 & Carl Hubbell & 2.94 & Phil Niekro & 3364 \\ 
  13 & Zack Greinke & 90.23 & Clayton Kershaw & 88.83 & Curt Schilling & 2.96 & Justin Verlander & 3297 \\ 
  14 & Gaylord Perry & 89.50 & Don Sutton & 82.98 & John Smoltz & 2.96 & Pedro Martinez & 3113 \\ 
  15 & Steve Carlton & 88.70 & Max Scherzer & 82.14 & Whitey Ford & 2.96 & John Smoltz & 3106 \\ 
  16 & Pedro Martinez & 87.20 & Pedro Martinez & 82.13 & Bob Gibson & 2.97 & Bob Feller & 3104 \\ 
  17 & Nolan Ryan & 86.85 & Zack Greinke & 80.28 & Jim Palmer & 2.97 & Fergie Jenkins & 3088 \\ 
  18 & Mike Mussina & 84.62 & Mike Mussina & 80.08 & Zack Greinke & 3.01 & Curt Schilling & 3036 \\ 
  19 & Curt Schilling & 82.09 & John Smoltz & 79.36 & Tim Hudson & 3.03 & Clayton Kershaw & 2980 \\ 
  20 & Tom Glavine & 81.89 & Pete Alexander & 77.55 & Juan Marichal & 3.05 & CC Sabathia & 2960 \\ 
  21 & Robin Roberts & 79.01 & Phil Niekro & 77.47 & Steve Carlton & 3.07 & Warren Spahn & 2955 \\ 
  22 & Fergie Jenkins & 77.83 & Curt Schilling & 77.00 & Tom Glavine & 3.10 & Zack Greinke & 2916 \\ 
  23 & Bob Gibson & 77.00 & Bob Gibson & 76.45 & Félix Hernández & 3.10 & Frank Tanana & 2849 \\ 
  24 & Roy Halladay & 76.07 & Fergie Jenkins & 75.97 & Kevin Brown & 3.12 & Bob Gibson & 2836 \\ 
  25 & CC Sabathia & 74.50 & Tommy John & 75.80 & Adam Wainwright & 3.12 & Lefty Grove & 2826 \\ 
   \hline
\end{tabular}
\caption{Top 25 pitching careers according to era-adjusted bWAR (ebWAR), era-adjusted fWAR (efWAR), era-adjusted earned run average (minimum 3000 adjusted inning pitched), and era-adjusted strikeouts.}
\label{Tab:ecareerpitchers}
\end{table}

\begin{table}[ht]
\scriptsize
\centering
\begin{tabular}{rlrlrlrlr}
  \hline
 & name & ebWAR & name & efWAR & name & ebJAWS & name & efJAWS \\ 
  \hline
1 & Barry Bonds & 153.89 & Barry Bonds & 145.24 & Barry Bonds & 109.14 & Roger Clemens & 103.54 \\ 
  2 & Roger Clemens & 145.88 & Roger Clemens & 141.25 & Roger Clemens & 107.47 & Barry Bonds & 103.21 \\ 
  3 & Willie Mays & 144.08 & Willie Mays & 135.39 & Willie Mays & 105.17 & Willie Mays & 98.72 \\ 
  4 & Babe Ruth & 137.98 & Henry Aaron & 128.05 & Babe Ruth & 100.70 & Babe Ruth & 90.29 \\ 
  5 & Henry Aaron & 135.60 & Greg Maddux & 120.73 & Henry Aaron & 95.11 & Henry Aaron & 89.52 \\ 
  6 & Alex Rodriguez & 120.29 & Babe Ruth & 120.28 & Alex Rodriguez & 91.66 & Greg Maddux & 88.51 \\ 
  7 & Stan Musial & 119.51 & Stan Musial & 113.03 & Stan Musial & 88.38 & Randy Johnson & 85.78 \\ 
  8 & Ty Cobb & 114.48 & Alex Rodriguez & 110.30 & Randy Johnson & 88.20 & Alex Rodriguez & 84.35 \\ 
  9 & Greg Maddux & 113.66 & Randy Johnson & 109.77 & Albert Pujols & 86.33 & Stan Musial & 83.61 \\ 
  10 & Albert Pujols & 111.86 & Ty Cobb & 108.77 & Greg Maddux & 85.60 & Ted Williams & 82.72 \\ 
  11 & Randy Johnson & 110.81 & Nolan Ryan & 108.30 & Mike Schmidt & 84.53 & Mike Schmidt & 82.20 \\ 
  12 & Mike Schmidt & 109.58 & Ted Williams & 107.75 & Lefty Grove & 84.52 & Lefty Grove & 79.31 \\ 
  13 & Rickey Henderson & 109.08 & Mike Schmidt & 106.41 & Ted Williams & 83.54 & Ty Cobb & 78.85 \\ 
  14 & Ted Williams & 107.86 & Rickey Henderson & 103.90 & Ty Cobb & 82.62 & Rickey Henderson & 78.83 \\ 
  15 & Tom Seaver & 104.31 & Bert Blyleven & 101.82 & Rickey Henderson & 82.14 & Steve Carlton & 78.32 \\ 
  16 & Lefty Grove & 102.54 & Steve Carlton & 100.34 & Justin Verlander & 80.34 & Nolan Ryan & 76.88 \\ 
  17 & Tris Speaker & 102.26 & Lefty Grove & 98.80 & Joe Morgan & 79.03 & Albert Pujols & 75.80 \\ 
  18 & Justin Verlander & 100.23 & Albert Pujols & 97.34 & Tom Seaver & 78.51 & Justin Verlander & 75.29 \\ 
  19 & Joe Morgan & 100.17 & Joe Morgan & 96.07 & Cal Ripken Jr & 77.54 & Joe Morgan & 75.13 \\ 
  20 & Frank Robinson & 99.93 & Frank Robinson & 95.92 & Mike Trout & 77.26 & Bert Blyleven & 75.12 \\ 
  21 & Mel Ott & 99.74 & Mel Ott & 95.72 & Rogers Hornsby & 76.62 & Rogers Hornsby & 74.28 \\ 
  22 & Bert Blyleven & 97.69 & Tris Speaker & 95.13 & Lou Gehrig & 75.70 & Mike Trout & 73.90 \\ 
  23 & Cal Ripken Jr & 97.39 & Justin Verlander & 95.07 & Wade Boggs & 75.61 & Mickey Mantle & 73.71 \\ 
  24 & Rogers Hornsby & 97.01 & Gaylord Perry & 94.45 & Clayton Kershaw & 75.12 & Cal Ripken Jr & 73.62 \\ 
  25 & Lou Gehrig & 95.87 & Rogers Hornsby & 94.42 & Mickey Mantle & 75.09 & Lou Gehrig & 73.41 \\ 
   \hline
\end{tabular}
\caption{Top 25 careers according to era-adjusted bWAR (ebWAR), era-adjusted fWAR (efWAR), era-adjusted JAWS computed using bWAR (ebJAWS), and era-adjusted JAWS computed using fWAR (efJAWS). JAWS is the average of a players' career WAR and the a players' total WAR from his seven best seasons.}
\label{Tab:ecareercombined}
\end{table}

\begin{table}[ht]
\centering
\begin{tabular}{rlcclcc}
  \hline
 & name & years & BA & name & years & AB per HR \\ 
  \hline
1 & Jose Altuve & 2014-2017 & 0.367 & Barry Bonds & 2001-2004 & 10.86 \\ 
  2 & Tony Gwynn & 1994-1997 & 0.366 & Mark McGwire & 1995-1998 & 11.15 \\ 
  3 & Rod Carew & 1974-1977 & 0.363 & Babe Ruth & 1918-1921 & 11.35 \\ 
  4 & Miguel Cabrera & 2010-2013 & 0.355 & Giancarlo Stanton & 2014-2017 & 11.85 \\ 
  5 & Wade Boggs & 1985-1988 & 0.353 & Albert Pujols & 2008-2011 & 12.20 \\ 
  6 & Ichiro Suzuki & 2001-2004 & 0.353 & Eddie Mathews & 1953-1956 & 12.34 \\ 
  7 & Barry Bonds & 2001-2004 & 0.352 & Willie Stargell & 1970-1973 & 12.45 \\ 
  8 & Joe Mauer & 2006-2009 & 0.350 & Jose Canseco & 1988-1991 & 12.46 \\ 
  9 & Roberto Clemente & 1964-1967 & 0.345 & Mike Schmidt & 1980-1983 & 12.57 \\ 
  10 & Joe DiMaggio & 1938-1941 & 0.345 & Jose Bautista & 2010-2013 & 12.68 \\ 
  11 & Albert Pujols & 2003-2006 & 0.343 & Gorman Thomas & 1978-1981 & 12.80 \\ 
  12 & Don Mattingly & 1984-1987 & 0.341 & Ralph Kiner & 1949-1952 & 12.87 \\ 
  13 & Mike Piazza & 1995-1998 & 0.341 & Khris Davis & 2015-2018 & 12.89 \\ 
  14 & Willie Mays & 1957-1960 & 0.340 & Aaron Judge & 2020-2023 & 13.25 \\ 
  15 & Matty Alou & 1966-1969 & 0.339 & Ted Williams & 1944-1947 & 13.26 \\ 
  16 & Tim Anderson & 2019-2022 & 0.338 & Sammy Sosa & 1998-2001 & 13.55 \\ 
  17 & Stan Musial & 1943-1946 & 0.338 & Frank Howard & 1967-1970 & 13.56 \\ 
  18 & Rogers Hornsby & 1922-1925 & 0.335 & Mickey Mantle & 1960-1963 & 13.66 \\ 
  19 & Ted Williams & 1943-1946 & 0.335 & David Ortiz & 2012-2015 & 13.71 \\ 
  20 & Ty Cobb & 1912-1915 & 0.334 & Nelson Cruz & 2017-2020 & 13.88 \\ 
  21 & Trea Turner & 2019-2022 & 0.334 & Jimmie Foxx & 1937-1940 & 13.91 \\ 
  22 & Henry Aaron & 1956-1959 & 0.333 & Carlos Pena & 2007-2010 & 13.91 \\ 
  23 & Cecil Cooper & 1980-1983 & 0.333 & Dave Kingman & 1976-1979 & 13.97 \\ 
  24 & Freddie Freeman & 2020-2023 & 0.333 & Darryl Strawberry & 1985-1988 & 13.98 \\ 
  25 & Nap Lajoie & 1901-1904 & 0.333 & Jim Thome & 2001-2004 & 14.01 \\ 
   \hline
\end{tabular}
\caption{Top 25 four-year peaks by batting average and at bats per home run (minimum 2000 era-adjusted plate appearances). }
\label{Tab:epeakbatting}
\end{table}

\begin{figure}[h]
\begin{center}
\includegraphics[scale = 0.40]{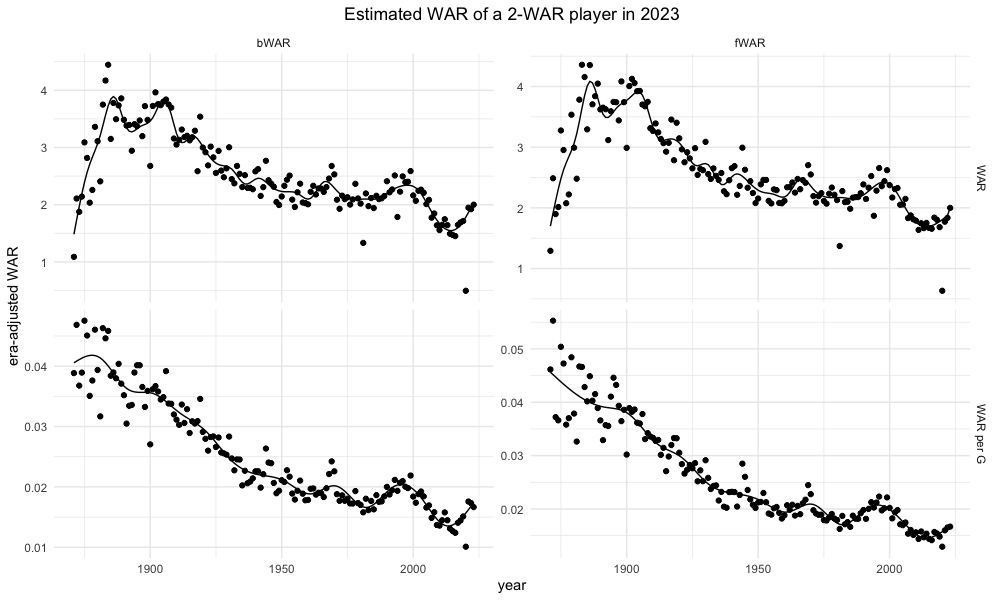}	
\end{center}	
\caption{The figure in the top left shows the estimated bWAR values over time corresponding to a hypothetical player with 2 bWAR in 2023; The figure in the top right shows the estimated fWAR values over time corresponding to a hypothetical player with 2 fWAR in 2023; The figure in the bottom left shows the estimated bWAR per game values over time corresponding to a hypothetical player with 2 bWAR in 2023; The figure in the bottom right shows the estimated fWAR values over time corresponding to a hypothetical player with 2 fWAR in 2023. Note that there were fewer games played in the early years of baseball.}
\label{Fig:WAR}
\end{figure}

Figure~\ref{Fig:WAR} displays estimates of the WAR that a hypothetical 2-WAR player in 2023 would produce in other seasons. Interestingly, this plot does not perfectly mirror the talent pool presented in Figure~\ref{Fig:talentpool}. In particular, a 2-WAR player in 2023 is estimated to barely exceed 2 WAR in the 1950s despite the talent pool being far larger in 2023. One reason for this is MLB expansion. There were, respectively, 16 and 30 teams in the 1950s and 2023\footnote{\url{https://en.wikipedia.org/wiki/Timeline_of_Major_League_Baseball}}. Fewer players were separating a 2-WAR player from the maximum achiever in the 1950s than in 2023. Thus a 2-WAR player in the 1950s would correspond to a higher quantile of the ordered sample of latent baseball talents than a 2-WAR player in 2023. This partially offsets the fact that the talent pool was larger in 2023 than it was in the 1950s.

\subsection{A closer look at Barry Bonds and Babe Ruth peak home run seasons}
\label{sec:BondsRuth}

Table~\ref{Tab:epeakbatting} shows that Barry Bonds and Babe Ruth, respectively, rank first and third in four-year peak home run rate. We unpack how our modeling works for the top three seasons by home run rate for each of these players. Notice that $\widetilde{F}_{Y_i}$ can be rewritten as
\begin{align*}
  \widetilde{F}_{Y_i}\left(Y_{i,(n_i)}\right) 
    &= \frac{n_i - 1}{n_i} + \frac{Y_{i,(n_i)} - \widetilde{Y}_{i,(n_i)}}{n_i\left(Y_{i,(n_i)} + Y_i^{\star\star} - \widetilde{Y}_{i,(n_i)}\right)} \\
    &= 1 - \frac{1}{n_i} + \frac{1}{n_i}\left(\frac{\frac{Y_{i,(n_i)} - Y_{i,(n_i-1)}}{2}}{\frac{Y_{i,(n_i)} - Y_{i,(n_i-1)}}{2} + Y_{i}^{\star\star}}\right).
\end{align*}
Thus, how far someone stands above their peers is a balance between how far above their closest peer and $Y_i^{\star\star}$ which is calculated in Section~\ref{Sec:ystar}. With this in mind, Table~\ref{Fig:RuthBonds} displays the results for Barry Bonds and Babe Ruth, where we define: 
\begin{itemize}
	\item $\text{diff} = Y_{i,(n_i)} - Y_{i,(n_i-1)}$, where these differences are defined after shrinkage is applied to the handedness park-factor adjusted home run rates. The shrinkage method that we used follows from \cite{SM13, SM16}. See the Supplementary Materials for how shrinkage is applied.
	\item $\text{balance} = \frac{Y_{i,(n_i)} - Y_{i,(n_i-1)}}{2}$.
	\item pBeta is calculated as $U_{i,(n_i)}$ in \eqref{eq:nonparest}.
	\item X is calculated is $X_{i,(N_i)}$ in \eqref{eq:nonparest}.
\end{itemize}
Table~\ref{Fig:RuthBonds} reveals that Babe Ruth stood further from his peers than Barry Bonds did, his diff, $Y^{\star\star}$, and pBeta values are all higher. But Barry Bonds ended up with higher estimated talent scores due to his playing during a time when the talent pool was much larger.

\begin{figure}
\includegraphics[scale=0.4]{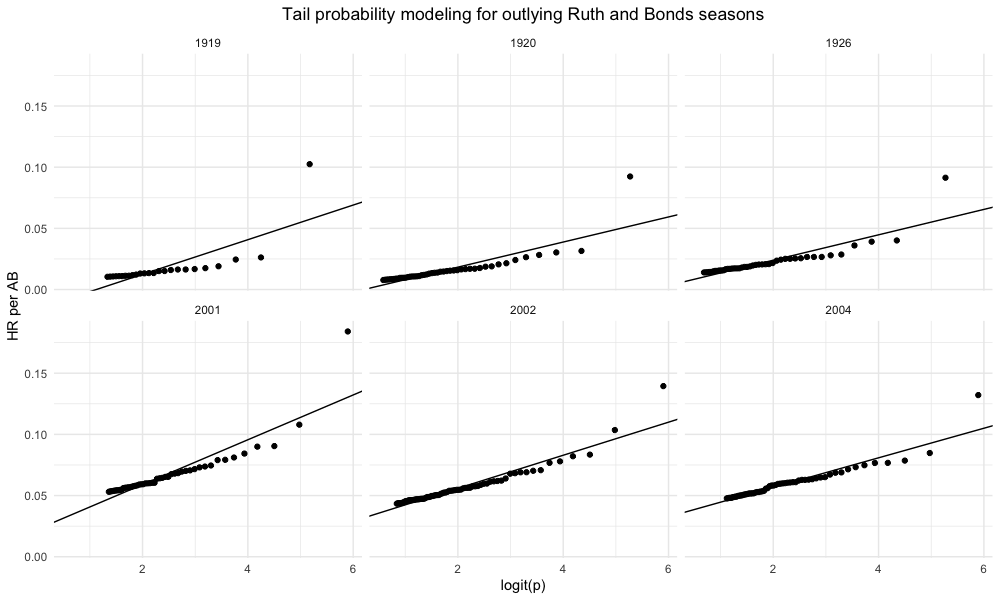}
\caption{Tail probability model fits for home run rates for the top three era-adjusted seasons by Babe Ruth and Barry Bonds according to home run rate.}
\label{Fig:tailprobs}
\end{figure}

Tail probability model fits are displayed in Figure~\ref{Fig:tailprobs}. Unsurprising,  both Barry Bonds and Babe Ruth's home rates are high leverage points with Ruth's 1919 and 1920 seasons having especially large influence on our tail-probability models.

\begin{table}[ht]
\begin{tabular}{lrccccccrr}
  \hline
name & year & AB per HR & diff & $Y^{\star\star}$ & balance & n & pBeta & N & X \\ 
  \hline
Babe Ruth & 1919 & 10.95 & 0.0662 & 0.00108 & 0.968 & 118 & 0.969 & 2020530 & 5355122 \\ 
Babe Ruth & 1920 & 10.88 & 0.0508 & 0.00040 & 0.985 & 130 & 0.985 & 2517182 & 12061976 \\ 
Babe Ruth & 1926 & 10.83 & 0.0413 & 0.00071 & 0.967 & 130 & 0.967 & 3499956 & 8272038 \\ 
Barry Bonds & 2001 & 10.86 & 0.0662 & 0.00140 & 0.959 & 243 & 0.960 & 11200119 & 18901295 \\ 
Barry Bonds & 2002 & 10.83 & 0.0259 & 0.00132 & 0.907 & 244 & 0.912 & 11725596 & 9660871 \\ 
Barry Bonds & 2004 & 10.76 & 0.0373 & 0.00161 & 0.920 & 242 & 0.924 & 12829045 & 11901386 \\ 
   \hline
\end{tabular}
\caption{Comparison of the top three era-adjusted seasons by Babe Ruth and Barry Bonds according to home run rate (AB per HR). $N$ is the size of the talent pool, $n$ is the number of full time players, $Y^{\star\star}$ is the upper bound calculated in Section~\ref{Sec:ystar}, and all other quantities are defined in Section~\ref{sec:BondsRuth}. Note that the absolute difference between the estimated talent scores in the $X$ appears large, but they are quite small when mapped to Pareto percentiles.}
\label{Fig:RuthBonds}
\end{table}

\section{Sensitivity Analyses}
\label{sensitivity}

In this section, we investigate how results change when we vary our implementation. Broadly speaking, we perform sensitivity analyses on three fronts: 1) changes to the talent pool; 2) investigations into relaxing the specification that the most talented people are in the MLB; 3) a multiverse analysis \citep{steegen2016increasing} in which we vary implementation specifications.

\subsection{Talent pool sensitivity}
\label{sec:talentpoolsensitivity}

We consider four alternative estimates of the talent pool. They are as follows: 
\begin{itemize}
	\item[A)] Our original estimate of the talent pool as described in Section~\ref{sec:MLBpop}.
	\item[B)] An estimate of the talent pool where interest in baseball is calculated with 0.75 weight to surveys in which people list baseball as their favorite sport and 0.25 weight to general baseball interest.
	\item[C)] An estimate of the talent pool where interest in baseball is calculated with 1 weight to surveys in which people list baseball as their favorite sport and 0 weight to general baseball interest.
	\item[D)] An estimate of the talent pool with 0.9 interest in baseball from 1871-1949, our original estimate of baseball interest post-1964, and a linear decrease in baseball interest from 0.9 in 1949 to our original estimated baseball interest in 1964.
	\item[E)] A talent pool that is fixed over time.
\end{itemize}
Talent pools B-E are constructed to elevate the standing of baseball players from the past relative to more modern players. To the best of our knowledge, surveys of baseball interest do not exist prior to 1937. In light of this, we made an intelligent guess about baseball interest prior to 1937 which is unchanged in talent pools B and C. Thus, the talent pool size of older era players is relatively larger in talent pools B and C than in our original talent pool. We can see the effect of these changes in Table~\ref{Tab:eJAWSsensitivity} which displays era-adjusted JAWS (the average of ebJAWS and efJAWS) ranking lists for each of the considered talent pool estimates. Talent pools B-C yield similar results as our original talent pool with slight gains made by older era players. The results yielded by talent pool D clearly favor older era players more than those yielded by talent pools A-C. Talent pool E is the most favorable to older era players with Babe Ruth distancing himself from the field, and Cap Anson appearing firmly in the top 10. Note that Cap Anson's eJAWS is elevated by an increase in games played in the number of games that he likely would play in our common mapping environment for comparing players (1977-1989 NL seasons with the 1981 strike-shortened season removed).

\begin{table}[ht]
\centering
\tiny
\begin{tabular}{rlrlrlrlrlr}
  \hline
& \multicolumn{2}{c}{A} & \multicolumn{2}{c}{B} & \multicolumn{2}{c}{C} & \multicolumn{2}{c}{D} & \multicolumn{2}{c}{E} \\
 & name & eJAWS & name & eJAWS & name & eJAWS & name & eJAWS & name & eJAWS \\ 
 \hline
1 & Barry Bonds & 106.17 & Barry Bonds & 106.47 & Willie Mays & 106.10 & Babe Ruth & 112.24 & Babe Ruth & 117.14 \\ 
  2 & Roger Clemens & 105.50 & Roger Clemens & 104.84 & Babe Ruth & 105.83 & Barry Bonds & 106.50 & Ty Cobb & 108.87 \\ 
  3 & Willie Mays & 101.94 & Willie Mays & 104.02 & Barry Bonds & 105.22 & Roger Clemens & 104.35 & Willie Mays & 107.51 \\ 
  4 & Babe Ruth & 95.50 & Babe Ruth & 100.28 & Roger Clemens & 103.36 & Willie Mays & 104.19 & Barry Bonds & 106.98 \\ 
  5 & Henry Aaron & 92.32 & Henry Aaron & 94.03 & Henry Aaron & 97.05 & Ty Cobb & 100.51 & Cy Young & 106.41 \\ 
  6 & Alex Rodriguez & 88.00 & Stan Musial & 88.33 & Ty Cobb & 92.31 & Walter Johnson & 95.29 & Roger Clemens & 105.75 \\ 
  7 & Greg Maddux & 87.06 & Alex Rodriguez & 87.46 & Stan Musial & 91.79 & Henry Aaron & 93.98 & Cap Anson & 105.38 \\ 
  8 & Randy Johnson & 86.99 & Greg Maddux & 86.28 & Lefty Grove & 89.54 & Stan Musial & 93.94 & Walter Johnson & 104.81 \\ 
  9 & Stan Musial & 86.00 & Randy Johnson & 86.07 & Ted Williams & 87.53 & Lefty Grove & 92.66 & Honus Wagner & 101.93 \\ 
  10 & Mike Schmidt & 83.37 & Ty Cobb & 86.04 & Alex Rodriguez & 85.84 & Honus Wagner & 91.15 & Henry Aaron & 98.54 \\ 
  11 & Ted Williams & 83.13 & Ted Williams & 85.27 & Walter Johnson & 85.74 & Tris Speaker & 90.48 & Stan Musial & 98.13 \\ 
  12 & Lefty Grove & 81.91 & Lefty Grove & 85.18 & Greg Maddux & 84.66 & Cy Young & 89.53 & Tris Speaker & 97.72 \\ 
  13 & Albert Pujols & 81.07 & Mike Schmidt & 83.93 & Mike Schmidt & 84.41 & Ted Williams & 89.32 & Lefty Grove & 93.98 \\ 
  14 & Ty Cobb & 80.74 & Rickey Henderson & 79.86 & Randy Johnson & 84.18 & Rogers Hornsby & 88.20 & Eddie Collins & 93.82 \\ 
  15 & Rickey Henderson & 80.48 & Albert Pujols & 79.37 & Rogers Hornsby & 83.75 & Alex Rodriguez & 87.94 & Ted Williams & 92.09 \\ 
  16 & Justin Verlander & 77.82 & Rogers Hornsby & 79.16 & Tris Speaker & 82.72 & Greg Maddux & 86.59 & Rogers Hornsby & 92.03 \\ 
  17 & Joe Morgan & 77.08 & Joe Morgan & 78.39 & Lou Gehrig & 81.15 & Randy Johnson & 85.56 & Nap Lajoie & 88.60 \\ 
  18 & Cal Ripken Jr & 75.58 & Walter Johnson & 77.96 & Joe Morgan & 79.84 & Eddie Collins & 85.07 & Greg Maddux & 87.34 \\ 
  19 & Mike Trout & 75.58 & Tris Speaker & 77.37 & Mel Ott & 79.75 & Mel Ott & 84.97 & Randy Johnson & 86.86 \\ 
  20 & Rogers Hornsby & 75.45 & Lou Gehrig & 77.31 & Honus Wagner & 79.60 & Lou Gehrig & 84.77 & Mel Ott & 86.76 \\ 
  21 & Bert Blyleven & 75.06 & Justin Verlander & 76.57 & Rickey Henderson & 78.68 & Mike Schmidt & 83.19 & Lou Gehrig & 86.27 \\ 
  22 & Lou Gehrig & 74.56 & Mickey Mantle & 76.25 & Mickey Mantle & 78.68 & Cap Anson & 81.33 & Alex Rodriguez & 85.12 \\ 
  23 & Mickey Mantle & 74.40 & Mike Trout & 75.43 & Tom Seaver & 76.67 & Albert Pujols & 81.03 & Pete Alexander & 83.86 \\ 
  24 & Steve Carlton & 74.40 & Bert Blyleven & 75.43 & Steve Carlton & 76.54 & Rickey Henderson & 80.28 & Mike Schmidt & 83.62 \\ 
  25 & Tom Seaver & 74.15 & Steve Carlton & 75.33 & Frank Robinson & 76.46 & Jimmie Foxx & 78.52 & Mickey Mantle & 83.50 \\ 
   \hline
   \end{tabular}
\caption{Era-adjusted JAWS (eJAWS) rankings computed with respect to talent pool estimates A-E. eJAWS is the average of era-adjusted JAWS computed using bWAR (ebJAWS), and era-adjusted JAWS computed using fWAR (efJAWS). JAWS is the average of a players' career WAR and the a players' total WAR from his seven best seasons.}
\label{Tab:eJAWSsensitivity}
\end{table}

\subsection{Specification that the most talented people are in the MLB}

In this section, we perform a sensitivity analysis on our assumption that the most talented players in the MLB. For example, it could be possible that our MLB interest adjustment may not capture dual-sport athletes such as Kyler Murray and Pat Mahomes who are currently playing in the NFL but might be among the most talented baseball players. 

In this sensitivity analysis, we will suppose that the 10th, 20th, ..., and 100th talented potential baseball players fail to start their sports career in baseball, which indicates the player with the 10th largest bWAR or fWAR is paired with the 11th largest talent, the player with 20th largest bWAR or fWAR is paired with 22nd largest talent, and so on. Then we mapped their talents into the common mapping environment we built before and computed the era-adjusted bWAR and fWAR. We perform this analysis for the seasons after the 1950 season corresponding to a potential effect due to the collapse of the Minor Leagues. We will also perform this analysis for the seasons after the 1994 season based on the effect of the MLB strike which saw the World Series get canceled. 

Table \ref{Tab:eJAWSremoved} shows the era-adjusted JAWS (eJAWS) rankings computed with respect to the 10th, 20th, ..., and 100th talented potential baseball players who fail to start their sports career in baseball. The ranking of the top 10 is stable. After the top 10, we do see modern players beginning to take a hit. For example, Albert Pujols ranks 13th in our original analysis, and he falls slightly when some people who may have had better seasons than him are removed from the talent pool. The effect of the sensitivity analysis is most apparent for Mike Trout and Justin Verlander who are active MLB players at the time this was written. These players drop out of the top 25 list completely when very talented people are completely removed from the pool.

\begin{table}[ht]
\centering
\begin{tabular}{rlrlrlr}
  \hline
 & \multicolumn{2}{c}{Original analysis} & \multicolumn{2}{c}{Remove after 1950} & 
 \multicolumn{2}{c}{Remove after 1994} \\
 & name & eJAWS & name & eJAWS & name & eJAWS \\ 
  \hline
1 & Barry Bonds & 106.17 & Barry Bonds & 105.90 & Barry Bonds & 106.03 \\ 
  2 & Roger Clemens & 105.50 & Roger Clemens & 104.96 & Roger Clemens & 105.15 \\ 
  3 & Willie Mays & 101.94 & Willie Mays & 101.72 & Willie Mays & 102.08 \\ 
  4 & Babe Ruth & 95.50 & Babe Ruth & 95.50 & Babe Ruth & 95.50 \\ 
  5 & Henry Aaron & 92.32 & Henry Aaron & 91.54 & Henry Aaron & 92.24 \\ 
  6 & Alex Rodriguez & 88.00 & Alex Rodriguez & 87.39 & Alex Rodriguez & 87.39 \\ 
  7 & Greg Maddux & 87.06 & Randy Johnson & 86.09 & Randy Johnson & 86.65 \\ 
  8 & Randy Johnson & 86.99 & Greg Maddux & 85.75 & Greg Maddux & 86.46 \\ 
  9 & Stan Musial & 86.00 & Stan Musial & 84.79 & Stan Musial & 85.85 \\ 
  10 & Mike Schmidt & 83.36 & Mike Schmidt & 83.05 & Mike Schmidt & 83.37 \\ 
  11 & Ted Williams & 83.13 & Ted Williams & 82.88 & Ted Williams & 83.18 \\ 
  12 & Lefty Grove & 81.91 & Lefty Grove & 81.91 & Lefty Grove & 81.91 \\ 
  13 & Albert Pujols & 81.06 & Ty Cobb & 80.72 & Ty Cobb & 80.72 \\ 
  14 & Ty Cobb & 80.74 & Rickey Henderson & 79.13 & Rickey Henderson & 79.38 \\ 
  15 & Rickey Henderson & 80.48 & Albert Pujols & 78.84 & Albert Pujols & 78.84 \\ 
  16 & Justin Verlander & 77.82 & Joe Morgan & 76.00 & Joe Morgan & 77.09 \\ 
  17 & Joe Morgan & 77.09 & Rogers Hornsby & 75.44 & Rogers Hornsby & 75.44 \\ 
  18 & Cal Ripken Jr & 75.58 & Lou Gehrig & 74.56 & Bert Blyleven & 75.06 \\ 
  19 & Mike Trout & 75.58 & Cal Ripken Jr & 74.02 & Cal Ripken Jr & 74.81 \\ 
  20 & Rogers Hornsby & 75.44 & Mickey Mantle & 73.84 & Lou Gehrig & 74.56 \\ 
  21 & Bert Blyleven & 75.06 & Bert Blyleven & 73.78 & Mickey Mantle & 74.41 \\ 
  22 & Lou Gehrig & 74.56 & Tom Seaver & 73.05 & Steve Carlton & 74.40 \\ 
  23 & Mickey Mantle & 74.40 & Steve Carlton & 72.97 & Tom Seaver & 74.13 \\ 
  24 & Steve Carlton & 74.40 & Wade Boggs & 72.65 & Wade Boggs & 73.05 \\ 
  25 & Tom Seaver & 74.15 & Mel Ott & 72.64 & Mel Ott & 72.64 \\ 
   \hline
\end{tabular}
\caption{Era-adjusted JAWS (eJAWS) rankings computed with respect to the 10th, 20th, ..., and 100th most talented potential baseball players not being in the MLB for every season after some stated season (1950 and 1994). eJAWS is the average of era-adjusted JAWS computed using bWAR (ebJAWS), and era-adjusted JAWS computed using fWAR (efJAWS). JAWS is the average of a player's career WAR and the player's total WAR from his seven best seasons. The original analysis is our top 25 JAWS list without removing players.}
\label{Tab:eJAWSremoved}
\end{table}

\newpage

\subsection{Multiverse Analyses}

We investigate how modeling choices can affect results through two comparisons using a multiverse analytical approach \cite{steegen2016increasing}. The first comparison is made with respect to the batting averages of 1997 Tony Gwynn and 1911 Ty Cobb. The second comparison is made with respect to the at-bats per home run of 2001 Barry Bonds and 1920 Babe Ruth.

Four modeling choices were considered: the park-factor effect, the effect of the talent pool, whether or not the distribution of the batting statistics is parametric or nonparametric (only relevant for batting average comparisons, the parametric distribution for batting averages is the normal distribution), and the number of full-time players. 

The modeling choices have a more pronounced impact on our batting average comparison than our AB per home run comparison. We also note that in this comparison of batting averages we see that the modeling choice used in our analysis (park-factor = YES; Talent Pool = A; Distribution = nonparametric; League Size = historical) is the most favorable configuration for 1997 Tony Gwynn relative to 1911 Ty Cobb. See the Supplementary Materials for more details.

\section{Summary and Discussion} 
\label{summary}

Michael Schell went to great lengths to compare batters across eras in two books, \cite{SM13} and \cite{SM16}. Motivated by \cite{Gould96}, Schell also considered the standard deviation as a proxy for measuring a changing talent pool. On page 58 in \cite{SM16}, he outlined some problems with his standard deviation approach and called for a more sophisticated statistical method to handle difficulties with the changing talent pool. He said: ``Someday we will need to abandon the use of the standard deviation as a talent pool adjustment altogether and search for another talent pool adjustment method, likely involving more difficult statistical methods than those used in this book." Our work answers Schell's call through the development of a novel statistical model which directly connects achievement to talent after a careful estimate of the size of the talent pool is supplied as a modeling input. 

Estimation the talent pool comes with some level of subjectivity, and different estimates of the talent pool can yield different results. This can be seen in Section~\ref{sec:talentpoolsensitivity} where we consider results based alternative talent pool estimates, labeled B-E. We provide some reasons for the shortcomings of each of these alternative talent pools.
Baseball interest as defined in B and C above yield an estimated talent pool that is not alignment with a talent pool constructed from Latin American countries (details are at the end of the linked report included in Section~\ref{sec:MLBpop}). Talent pool D corresponds to the erosion of the Minor Leagues following their peak in 1949\footnote{\url{https://www.milb.com/milb/history/general-history}}. It is speculated that this collapse of the Minor Leagues led to an increase in MLB competitive balance because potentially exceptional players were lost due to professional baseball opportunities being removed (see the last paragraph of the Discussion in \cite{horowitz1997increasing}). Furthermore, it is occasionally speculated that this collapse was during a time in which baseball was thought of as an avenue for elevating the social status of lower-income men. However, the Minor League teams that were lost were lower quality teams, a majority of which were not affiliated with MLB farm systems \citep{land1994organizing}. \cite{sullivan1990minors} attributed the decline in the minor leagues to relocation of MLB teams to Minor League markets and the spread of television broadcasting games. \cite{bellamy2004did} noted that television went from 0.4 percent of U.S. households in 1948 to 87.1 percent in 1960. However, these authors concluded that ``the extension of MLB radio broadcasts, both national and local, in combination with a market correction, probably hurt the Minor Leagues, particularly in the first half of the 1950s, much more than the televised broadcasts that were in their infancy." \cite{riess1980professional} studied professional baseball and social mobility in the Progressive era. He concluded that professional baseball was ``greatly overrated as a source of upward mobility when the game was at its unchallenged height." In aggregate, we argue that our original talent pool estimate is more sensible than talent pool D, but we understand if the reader disagrees. Talent pool E is constant over time and therefore ignores the history of westward and southern expansion of baseball, integration, globalization, and effects of war, etc. Estimation of the talent pool is a research area of its own. Changing player salaries, changing media environments, demographic-specific interest in baseball, and the mechanisms by which international talent joins the MLB could all be avenues for improving the estimate of the talent pool. 


Importantly, we found that several great non-white and non-American players sit atop the era-adjusted WAR rankings. This is due to increased inclusion and general population increases in the talent pool over time. Our analyses found that more modern players such as Barry Bonds and Willie Mays have better careers than the legendary Babe Ruth. Current players to the time of this writing are adding legendary seasons to baseball's storied history. In particular, we highlight Shohei Ohtani and Aaron Judge's 2022 campaign, and Mike Trout's recent stretch of great seasons. All-time great baseball is being played right now, and we are excited for what the future holds. We are confident that this will remain true at the time you are reading this paper.

\section*{Supplementary Materials}

This article is accompanied by an extensive set of Supplementary Materials. First, there is a traditional supplement that complements this submission. There is also a GitHub repository containing further analyses and reports. This repository can be viewed here: 
\begin{center}
  \url{https://github.com/ecklab/era-adjustment-app-supplement}.	
\end{center}
We have also created a website that contains era-adjusted statistics for all qualifying baseball players from 1871-2023. This website includes a top 100 list by career ebWAR and efWAR as well as player pages that include era-adjusted and unadjusted statistics. This website can be viewed here:
\begin{center}
  \url{https://eckeraadjustment.web.illinois.edu/}	
\end{center}

\section*{Acknowledgments}

We are also grateful to 
Mark Armour,
Forrest W. Crawford, 
David Dalpiaz, 
Adam Darowski, 
Louis Moore, 
Ben Sherwood, 
Sihai Dave Zhao,
and many others for helpful comments and great discussions.

\section*{Supplementary Materials for Comparing baseball players across Eras via the Novel Full House Model}

This Supplementary Material includes: 
\begin{itemize}
	\item Pre-processing steps that were made for batting and pitching statistics.
	\item An investigation into the often-referenced claim that batting averages for full-time players are normally distributed.
	\item Additional sensitivity analyses for modeling assumptions.
	\item A visualization of the common-mapping environment used to compute era-adjusted bWAR for batters.
	\item Multiverse Analyses comparing the batting averages of 1997 Tony Gwynn and 1911 Ty Cobb, the at-bats per home run of 2001 Barry Bonds and 1920 Babe Ruth, top 25 career batting averages and top 25 four-year peaks by batting average.
	\item Brief theoretical details of $\widetilde{F}_Y(t)$, as defined in equation (3) in the manuscript.	
\end{itemize}


\vspace{12pt}
More materials can be seen on our GitHub repository:
\begin{center}
	\url{https://github.com/ecklab/era-adjustment-app-supplement}
\end{center}

This GitHub repository contains a detailed write-up on our estimation of the talent pool, codes for scrapping data from Baseball-Reference and Fangraphs, codes for the pre-process of batting and pitching stats, and a reproducible technical report that reproduces important tables and figures. 

\section{Pre-prossesing of batting and pitching statistics}

\subsubsection{Batting statistics}

In this section, we detail the considerations made for the batting statistics that we era-adjust using our methodology: batting average (BA), hits (H), home runs (HR), walks (BB), on-base percentage (OBP), bWAR, and fWAR. These statistics are all modeled on a rate basis where count statistics are converted to rates: we model HR/AB, BB/PA, bWAR/G, and fWAR/G.  We also adjust at-bats (AB) and plate appearances (PA) to accommodate changing season lengths throughout baseball's history. Handedness-specific park factor adjustments are also considered for BA and HR in our model. We apply the adjusted park index from \cite{SM16} to all ballparks from 1871-2023. We use nonparametric methods to model all statistics since these statistics (with the possible exception of BA) have not been demonstrated to follow any common distribution that we know. 

Our modeling will only be applied to full-time batters. We define full-time batters as anyone above the median number of PAs after screening out hitters who appeared in fewer than 75 PAs. This criterion is flexible enough to account for changing the number of games played over time as well as seasons shortened by labor strikes and pandemics. To deal with extreme statistics from the small sample size, we employed a shrinkage method that adjusted the raw statistics toward a global average. Our shrinkage method follows a shrinkage of team ballpark effect estimates motivated from \cite{SM16}. The \cite{SM16} method involved the weighted average of raw components (hits, home runs, etc.) for total team statistics of the form 
$$
\textrm{adjusted statistic} = \frac{\textrm{raw statistic} \times \textrm{total ABs} + 4000 \times \textrm{league average of the statistic}}{(\textrm{total ABs + 4000})}. 
$$
This shrinkage method from \cite{SM16} will shrink the team average to the league average. In our context of shrinkage for individual player statistics, we change the shrinkage factor $\frac{4000}{\textrm{totals AB + 4000}}$. For example, our shrinkage factor for bWAR per game is $\frac{7}{\textrm{individual games + 7}}$. The ratio $\frac{7}{\textrm{individual games + 7}}$ is approximately equal to $\frac{4000}{\textrm{totals AB + 4000}}$. Thus, 
$$
\textrm{adjusted bWAR per game} = \frac{\textrm{raw bWAR per game} \times \textrm{individual games} + 7 \times \textrm{league average bWAR per game}}{\textrm{individual games + 7}}. 
$$

We calculate mapped games and plate appearances by applying quantile mapping. Quantile mapping is based on the idea that a $p$th percentile player's games in one year are equal to a $p$th percentile player's games in another year. AB and PA also change across baseball history as the number of games and walk rates change. We calculate adjusted AB as:
 \begin{equation} \label{adjAB}
	\textrm{adjusted AB} =\textrm{adjusted PA - adjusted BB - HBP - SH - SF},
\end{equation}
where HBP, SH, and SF are short for, respectively, hit by pitch, sacrifice hits, and sacrifice flies, and adjusted PA is a player's quantile-mapped PA total from the season under study to that in the common-mapping environment (NL seasons from 1977 through 1989, excluding the 1981 strike-shortened season). 

From here, era-adjusted hits, walks, and home runs can be computed from the per AB rates and \eqref{adjAB} and adjusted PAs. For example, era-adjusted hits is equal to era-adjusted batting average multiplied by adjusted AB. Era-adjusted bWAR (ebWAR) or fWAR (efWAR) are obtained by multiplying era-adjusted bWAR per game or fWAR per game with adjusted games. We also compute era-adjusted OBP as
\begin{equation}\label{adjOBP}
\textrm{era-adjusted OBP} =\frac{\textrm{era-adjusted BA * adjusted AB + era-adjusted BB + HBP}}{\textrm{adjusted AB + era-adjusted BB + HBP + SF}}.
\end{equation}

We find that the Full House Model can harshly punish the tails of players' careers, especially for older-era players. This is due to players reaching or staying in the MLB during a less talented era of baseball history, and having these seasons translates to terrible play in the common context that we judge all players. For example, a late-career decline in the 1910s would correspond to a player who would likely be out of the MLB in the 1980s. We eliminate players who had an era-adjusted bWAR (ebWAR) or era-adjusted fWAR (efWAR) below the replacement level for more than half of their career seasons. We also set three rules to eliminate some players' poor early and late career seasons. The three rules are: 1) in at least 2 consecutive seasons, the ebWAR is below -1.5; 2) in at least 2 consecutive seasons, the efWAR is below -1.5; 3) in at least two consecutive seasons, no more than one ebWAR or efWAR can be more than 0.2. The value 0.2 is calculated from the average ebWAR or efWAR of the players that disappeared from the MLB from the 1977 season to the 1989 season with the exception of the 1981 strike-shortened season.

\subsubsection{Pitching statistics}
\label{pitching}

In this section, we detail the considerations made for the pitching statistics that we era-adjust using our methodology: earned-run average (ERA), strikeouts (SO), bWAR, and fWAR. We also adjust innings pitched (IP) using a similar quantile mapping approach that we applied to games or PA for batters. We will use nonparametric methods to measure these pitching statistics since these statistics have not been demonstrated to follow any common distribution we know. Note that smaller values of ERA are better, so we apply the Full House Model to the negation of ERA. Career trimming and shrinkage were applied to era-adjusted pitching statistics in the same way as to batting statistics.

Our modeling will only be applied to full-time pitchers. We define full-time pitchers as the $n$ pitchers who pitched the most innings above a cutoff. We now describe how the cutoff is defined: we get the average rotation size for each time in the MLB, and then we add these rotation sizes to arrive at $n$ full-time pitchers.

\section{Investigation of normality of batting averages} 
\label{batting}

\begin{figure}[h]
\centering
\includegraphics[scale = 0.25]{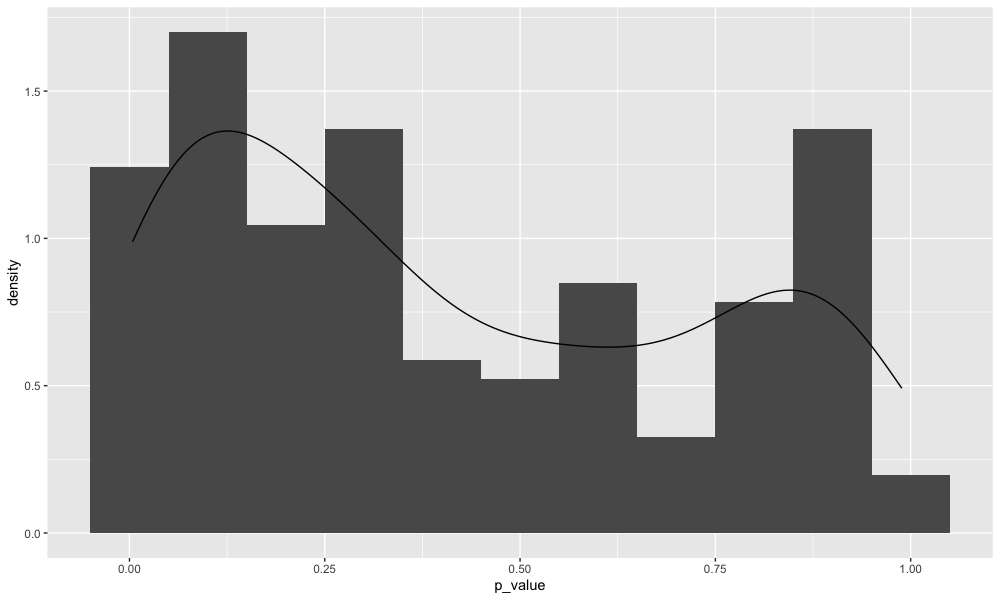}
\caption{Histogram of the p-values from Shapiro-Wilk test of normality on the BA distribution in each season.}
\label{Fig:p-value}
\end{figure}

We initially tried parametric distributions to model  BA since it is widely recognized that BA follows a normal distribution \citep{Gould96}. We perform a Shapiro-Wilk test \citep{SS65} of normality on the BA distribution for each season. Table \ref{Tab:normality} shows the p-values from the Shapiro-Wilk test for each season. Figure \ref{Fig:p-value} shows the histogram of the p-values from the Shapiro-Wilk test of normality on the BA distribution in each season. We observe that rejection rates from Shaprio-Wilk tests exceed nominal thresholds. Thus we will utilize the nonparametric distribution to measure the BA in each season because the normality of the BA distribution is not consistent with the data. 

\begin{table}[h]
\centering
\scalebox{1}{
\begin{tabular}{lc}
  \hline
  threshold & proportion of seasons exceeding threshold \\
  \hline
  0.05 & 0.15 \\
  0.1 & 0.24 \\
  0.2 & 0.34 \\
  0.3 & 0.39 \\
  \hline
\end{tabular}
}
\caption{The table shows the percent of seasons in which BA failed the Shapiro-Wilk test based on different p-value thresholds. }
\label{Tab:normality}
\end{table}

\section{Sensitivity analyses} 
\label{Sec:validation}

In this section, we validate our model to ensure the appropriateness of the assumption that the talent-generating process is Pareto distribution with parameter $\alpha = 1.16$ via a sensitivity analysis simulation. The goal of this analysis is to determine how many of the top 25 talented players by BA our method can correctly identify under a variety of simulation configurations, some of which are chosen to stretch the credibility of our method. 

In each simulation, we first randomly generate samples from four different talent generation distributions, which are the Pareto distribution with parameter $\alpha = 1.16$ (i.e. the talent generating process is correctly specified), Pareto distribution with parameter $\alpha = 3$, folded normal distribution with parameters $\mu = 0, \sigma = 1$, and standard normal distribution. Within each talent generation distribution, we vary the talent pool sizes $N_i$ for five different hypothetical leagues. Details of the simulation information are in Table \ref{simuinfo}. 

\begin{table}[ht!]
\centering
\scalebox{0.7}{
\begin{tabular}{c|c|c|c|c|c}
\hline
league & sample size & $\mu$ & $\sigma$ & improved estimation & deteriorated estimation\\ 
\hline
1 & $1*10^6$ & 0.280 & 0.040 & $0.5*10^6$ & $1*10^6$\\
2 & $2*10^6$ & 0.275 & 0.0375 & $1.5*10^6$ & $1*10^6$\\
3 & $4*10^6$ & 0.270 & 0.035 & $3.33*10^6$ & $1.33*10^6$\\
4 & $8*10^6$ & 0.265 & 0.0325 & $7*10^6$ & $2*10^6$\\
5 & $16*10^6$ & 0.260 & 0.030 & $14.4*10^6$ & $3.2*10^6$\\
\hline
\end{tabular}
}
\caption{Simulation study configurations. The sample size represents the talent pool for each of the five leagues. $\mu$ and $\sigma$ are the parameters of the normal distribution for generated batting averages. The improved estimation column is the estimated talent pool where the estimation improves. The deteriorated estimation column is the estimated talent pool where the estimation deteriorates. }
\label{simuinfo}
\end{table}

Then we select 300 people with the largest talents in each dataset and consider them as the full-time batters in MLB. Therefore, $n_i = 300$ for all $i$. We will generate BA from the talent scores using a normal distribution with parameters  Table~\ref{simuinfo}. For each league $i$, these values are generated as 
$
  Y_{i,(j)} = F_{Y_i}^{-1}\left(F^{-1}_{U_{i,(j)}}\left(F_{X_{i,(N_i-n_i+j)}}(X_{i,N_i-n_i+j})\right)|\theta_i\right),
$
where $\theta_i$ are the parameters for the normal distribution, and our choices for these parameters reflect the shrinking BA variability that has been observed over time \citep{Gould96}. 

We apply the Full House Model with $F_X$ assumed to be Pareto with $\alpha = 1.16$ to this generated data and investigate how well the Full House Model correctly identifies the top 25 players as judged by talent scores. We also consider misspecification in the estimation of the talent pool size.
 
We consider two underestimations of the talent pool where 1) the estimation improves as time increases and 2) the underestimation deteriorates as time increases. This configuration was chosen to be deliberately antagonistic to our method, especially when the talent-generating process was misspecified. The details of the information are in Table \ref{simuinfo}. 
We also compare our Full House Model to rankings based on Z-scores and unadjusted BA. Z-scores are a building block for the Power Transformation Method method.

\begin{figure}[ht]
\includegraphics[scale = 0.39]{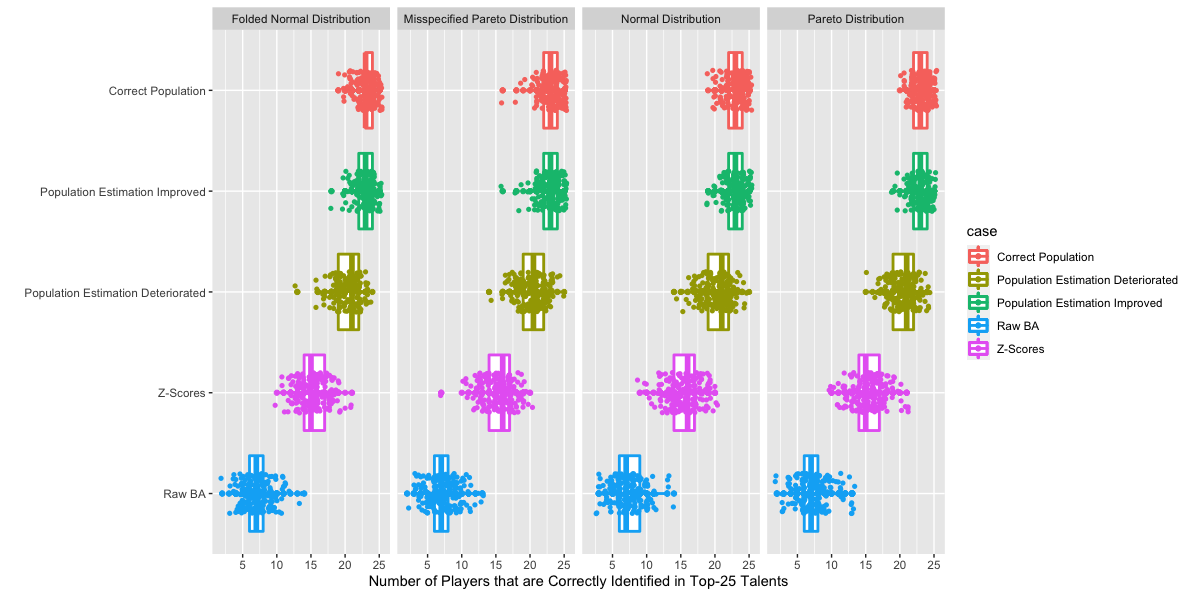}
\centering
\caption{Monte Carlo simulation investigating the number of correctly identified talents in a top 25 list under four different latent talent distribution $F_X$. Each column of the plot indicates a different talent generation distribution. The first three rows of the plot are variants of our Full House Model where $F_X$ is assumed to be Pareto with $\alpha = 1.16$. The first row displays the performance of our method with the talent pool correctly estimated. The second and third rows display the performance of our method with the talent pool incorrectly estimated. In the second row, we underestimate the talent pool where the estimation improves as the talent pool increases. In the third row, we underestimate the talent pool where the estimation deteriorates as the talent pool increases. The fourth row displays the performance of Z-scores and the fifth row displays the performance of raw unadjusted batting averages.}
\label{Fig:validation}
\end{figure}

200 Monte Carlo iterations of our simulation are performed and the results are depicted in Figure~\ref{Fig:validation} and Table~\ref{Tab:compare}. 

What we found is that our Pareto assumption with $\alpha = 1.16$ holds up well even when the talent-generating process and the talent pool is not correctly specified. Moreover, our method correctly identifies more top-25 talented players than both Z-scores and raw unadjusted batting averages. 

That being said, there is considerable overlap between the box plot in the second row and the box plot in the fourth row. This suggests that Z-scores may not be strictly worse than our method under misspecification. A more detailed look shows that this is not the case. During each simulation, we directly compare our method when the underestimation estimation of the talent pool deteriorates and Z-scores (third and fourth rows of Figure~\ref{Fig:validation}). Then we calculate the proportion of simulations that our model with the population estimation deteriorates strictly beats the Z-scores method and either beats or ties the Z-scores method. Table \ref{Tab:compare} indicates that our Full House Model is almost strictly better than the Z-scores method in each simulation. Thus, the Full House Model performs better than Z-scores even when $F_X$ and the estimated eligible population sizes are badly misspecified. 

\begin{table}[h]
\centering
\scalebox{0.7}{
\begin{tabular}{c|cccc}
  \hline
 & Correct Pareto distribution & Incorrect Pareto distribution & Normal distribution & Folded normal distribution \\ 
  \hline
beats or ties & 1 & 1 & 1 & 1 \\
strictly beats & 1 & 1 & 0.995 & 1 \\
   \hline
\end{tabular}
}
\caption{Comparison of results between rows 3 and 4 of Figure~\ref{Fig:validation}. The columns display the talent-generating distribution. The entries of the table display the proportion of Monte Carlo iterations for which the Full House Model with assumed Pareto talent generating process with $\alpha = 1.16$ and the talent pool estimated via the deteriorated estimation regime (see Table~\ref{simuinfo}) strictly beats or ties $Z$-scores in the number of top 25 talents correctly identified. }
\label{Tab:compare}
\end{table}

We validate the stability of rankings when we change the latent distribution $F_X$ to a folded normal distribution, the Pareto distribution with $\alpha = 3$, and the standard normal distribution. Table ~\ref{Tab:robustrankings} shows the rankings with these three different latent talent distributions. As expected, the distribution assumed on $F_X$ does not influence the results.

\begin{table}[t]
\centering
\scalebox{0.8}{
\begin{tabular}{lllllllll}
  \hline
 & \multicolumn{2}{c}{Standard normal} & \multicolumn{2}{c}{Folded normal ($\mu$ = 0, $\sigma$ = 1)} & \multicolumn{2}{c}{Pareto with $\alpha$ = 3} & \multicolumn{2}{c}{Pareto with $\alpha$ = 1.16} \\
 & name & ebWAR & name & ebWAR & name & ebWAR & name & ebWAR \\ 
  \hline
1 & Barry Bonds & 153.9 & Barry Bonds & 153.9 & Barry Bonds & 153.9 & Barry Bonds & 153.9 \\ 
  2 & Roger Clemens & 145.91 & Roger Clemens & 145.91 & Roger Clemens & 145.91 & Roger Clemens & 145.91 \\ 
  3 & Willie Mays & 144.1 & Willie Mays & 144.1 & Willie Mays & 144.1 & Willie Mays & 144.1 \\ 
  4 & Henry Aaron & 135.6 & Henry Aaron & 135.6 & Henry Aaron & 135.6 & Henry Aaron & 135.6 \\ 
  5 & Babe Ruth & 132.7 & Babe Ruth & 132.7 & Babe Ruth & 132.7 & Babe Ruth & 132.7 \\ 
  6 & Stan Musial & 119.51 & Stan Musial & 119.51 & Stan Musial & 119.51 & Stan Musial & 119.51 \\ 
  7 & Alex Rodriguez & 119.07 & Alex Rodriguez & 119.07 & Alex Rodriguez & 119.07 & Alex Rodriguez & 119.07 \\ 
  8 & Greg Maddux & 113.67 & Greg Maddux & 113.67 & Greg Maddux & 113.67 & Greg Maddux & 113.67 \\ 
  9 & Ty Cobb & 112 & Ty Cobb & 112 & Ty Cobb & 112 & Ty Cobb & 112 \\ 
  10 & Albert Pujols & 111.85 & Albert Pujols & 111.85 & Albert Pujols & 111.85 & Albert Pujols & 111.85 \\ 
  11 & Randy Johnson & 110.81 & Randy Johnson & 110.81 & Randy Johnson & 110.81 & Randy Johnson & 110.81 \\ 
  12 & Mike Schmidt & 109.59 & Mike Schmidt & 109.59 & Mike Schmidt & 109.59 & Mike Schmidt & 109.59 \\ 
  13 & Rickey Henderson & 109.06 & Rickey Henderson & 109.06 & Rickey Henderson & 109.06 & Rickey Henderson & 109.06 \\ 
  14 & Ted Williams & 108.04 & Ted Williams & 108.04 & Ted Williams & 108.04 & Ted Williams & 108.04 \\ 
  15 & Tom Seaver & 104.31 & Tom Seaver & 104.31 & Tom Seaver & 104.31 & Tom Seaver & 104.31 \\ 
  16 & Lefty Grove & 101.58 & Lefty Grove & 101.58 & Lefty Grove & 101.58 & Lefty Grove & 101.58 \\ 
  17 & Tris Speaker & 100.7 & Tris Speaker & 100.7 & Tris Speaker & 100.7 & Tris Speaker & 100.7 \\ 
  18 & Justin Verlander & 100.24 & Justin Verlander & 100.24 & Justin Verlander & 100.24 & Justin Verlander & 100.24 \\ 
  19 & Joe Morgan & 100.17 & Joe Morgan & 100.17 & Joe Morgan & 100.17 & Joe Morgan & 100.17 \\ 
  20 & Frank Robinson & 99.91 & Frank Robinson & 99.91 & Frank Robinson & 99.91 & Frank Robinson & 99.91 \\ 
  21 & Bert Blyleven & 97.69 & Bert Blyleven & 97.69 & Bert Blyleven & 97.69 & Bert Blyleven & 97.69 \\ 
  22 & Cal Ripken Jr & 97.41 & Cal Ripken Jr & 97.41 & Cal Ripken Jr & 97.41 & Cal Ripken Jr & 97.41 \\ 
  23 & Mel Ott & 96.72 & Mel Ott & 96.72 & Mel Ott & 96.72 & Mel Ott & 96.72 \\ 
  24 & Rogers Hornsby & 95.76 & Rogers Hornsby & 95.76 & Rogers Hornsby & 95.76 & Rogers Hornsby & 95.76 \\ 
  25 & Lou Gehrig & 95.68 & Lou Gehrig & 95.68 & Lou Gehrig & 95.68 & Lou Gehrig & 95.68 \\ 
   \hline
\end{tabular}
}
\caption{ebWAR rankings from Full House Model using different latent talent distributions: Standard normal, Folded normal ($\mu$ = 0, $\sigma$ = 1), Pareto with $\alpha$ = 3, and Pareto with $\alpha$ = 1.16. }
\label{Tab:robustrankings}
\end{table}

\section{Isotonic Regression for Common-Mapping Environment}

\begin{figure}[t]
\includegraphics[scale = 0.39]{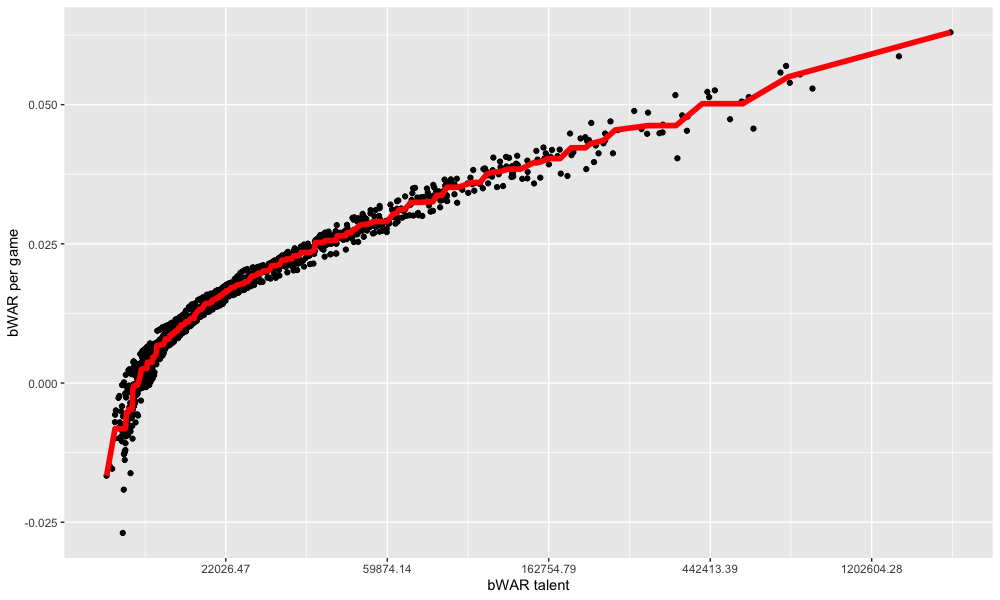}
\centering
\caption{The relationship between bWAR talent and bWAR per game. The black dots represent the observations from the full-time batters from the 1977 season to the 1989 season except for the 1981 strike-shortened season. The red dots represent the observations from the common mapping environment that we define in the manuscript. The talent values are computed using displayed equation (4) in the manuscript where $y^{\star\star}$ is computed in displayed equation (5) in the manuscript.}
\label{Fig:CME}
\end{figure}

Figure~\ref{Fig:CME} shows the relationship between bWAR talent and bWAR per game. The black dots represent the observations from the full-time batters from the 1977 season to the 1989 season except for the 1981 strike-shortened season. The red dots represent the observations from the common mapping environment that we define in the manuscript. Based on the figure above, the observations from the common mapping environment accurately depict the relationship between bWAR talent and bWAR per game from the 1977 season to the 1989 season except for the 1981 strike-shortened season.

\section{Pairings of the maximum talent scores with their corresponding era-adjusted bWAR for players in the common mapping environment}

\begin{figure}[t]
\includegraphics[scale = 0.39]{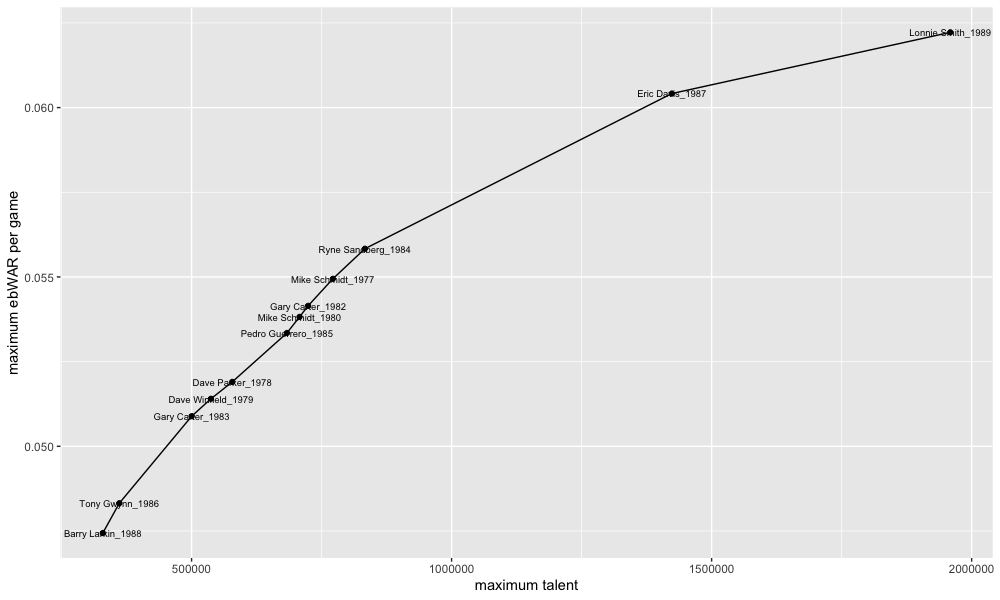}
\centering
\caption{The pairings of the maximum talent scores with their corresponding era-adjusted bWAR for players in the common mapping environment. The X-axis is their bWAR talent and the y-axis is their corresponding era-adjusted bWAR. }
\label{Fig:CME2}
\end{figure}

Figure~\ref{Fig:CME2} shows the pairings of the maximum talent scores with their corresponding era-adjusted bWAR for players in the common mapping environment. The X-axis is their bWAR talent and the Y-axis is their corresponding era-adjusted bWAR. Note that the the $y^{\star\star}$ value for bWAR per game in the common-mapping environment is 0.00325, and that Lonnie Smith's 1989 season ranked 116th overall.

\section{Multiverse Analyses}

In this section, we will show how modeling choices can affect results through two comparisons. The first comparison is made with respect to the batting averages of 1997 Tony Gwynn and 1911 Ty Cobb. The second comparison is made with respect to the at-bats per home run of 2001 Barry Bonds and 1920 Babe Ruth.

The four modeling choices considered are the park-factor effect, the talent pool, whether or not the distribution of the batting statistics is parametric or nonparametric (only relevant for batting average comparisons, the parametric distribution for batting averages is the normal distribution), and the number of full-time players. 

Results are displayed in Table~\ref{Tab:multiverseBAseason} and Table~\ref{Tab:multiverseHRseason}. The modeling choices have a more pronounced impact on the batting average comparison than the AB per home run comparison. When comparing batting averages we see that the modeling choice used in our analysis (park-factor = YES; Talent Pool = A; Distribution = nonparametric; League Size = historical) is the most favorable configuration for 1997 Tony Gwynn relative to 1911 Ty Cobb. We argue for talent pool A in the manuscript, use a handedness-specific park-factor method used by \cite{SM16}, and have demonstrated in this supplement that the normal distribution might be inappropriate as a model for batting averages. We think our modeling choices are sensible, but understand if the reader disagrees.

\begin{table}[ht]
\centering
\scalebox{1}{
\begin{tabular}{lllll}
  \hline
  Park factor & Talent pool & Distribution & League size & Difference\\
  \hline
  YES & A & parametric & historical & 0.020 \\
  & & & fixed & 0.008 \\
  & & nonparametric & historical & 0.040 \\
  & & & fixed & 0.040 \\
  & B & parametric & historical & 0.017 \\
  & & & fixed & 0.006 \\
  & & nonparametric & historical & 0.035 \\
  & & & fixed & 0.035 \\
  & C & parametric & historical & 0.014 \\
  & & & fixed & 0.003 \\
  & & nonparametric & historical & 0.023 \\
  & & & fixed & 0.023 \\
  & D & parametric & historical & 0.010 \\
  & & & fixed & -0.001 \\
  & & nonparametric & historical & 0.006 \\
  & & & fixed & 0.006 \\
  & E & parametric & historical & 0.006 \\
  & & & fixed & -0.005 \\
  & & nonparametric & historical & -0.002 \\
  & & & fixed & -0.002 \\
  NO & A & parametric & historical & 0.005 \\
  & & & fixed & 0.002 \\
  & & nonparametric & historical & 0.029 \\
  & & & fixed & 0.029 \\
  & B & parametric & historical & 0.003 \\
  & & & fixed & -0.001 \\
  & & nonparametric & historical & 0.027 \\
  & & & fixed & 0.027 \\
  & C & parametric & historical & 0.000 \\
  & & & fixed & -0.003 \\
  & & nonparametric & historical & 0.015 \\
  & & & fixed & 0.015 \\
  & D & parametric & historical & -0.004 \\
  & & & fixed & -0.008 \\
  & & nonparametric & historical & 0.007 \\
  & & & fixed & 0.007 \\
  & E & parametric & historical & -0.007 \\
  & & & fixed & -0.010 \\
  & & nonparametric & historical & -0.003 \\
  & & & fixed & -0.003 \\
  \hline
\end{tabular}
}
\caption{The table shows the value of the era-adjusted BA of Tony Gwynn in the 1997 season minus the era-adjusted BA of Ty Cobb in the 1911 season under different configurations. The Park factor column indicates whether we apply the park-factor adjustment to the BA. The Talent pool column indicates the population changes we apply to the talent pool. Population A - E are constructed according to the five different estimates of the talent pool in Section 5 of the manuscript. The Distribution column indicates we use parametric distribution and nonparametric distribution to measure the BA in each season. The League size column indicates how full-time players are calculated. Historical means that we use the number of full-time players observed for the given. Fixed means that the number of full-time players is set to the maximum number of seasonal full-time players observed in our data set. The Difference indicates the value of the BA of Tony Gwynn in the 1997 season minus the BA of Ty Cobb in the 1911 season under different configurations.}
\label{Tab:multiverseBAseason}
\end{table}

\begin{table}[ht]
\centering
\begin{tabular}{rlllr}
  \hline
 Park factor & Talent pool & League size & Difference \\ 
  \hline
YES & A & historical & -0.037 \\ 
    &  & fixed & -0.037 \\ 
    & B & historical & -0.012 \\ 
    &  & fixed & -0.012\\ 
    & C & historical & 0.007 \\ 
    &  & fixed & 0.007 \\ 
    & D & historical & 0.020 \\ 
    &  & fixed & 0.020 \\ 
    & E & historical & 0.028 \\ 
    &  & fixed & 0.028 \\ 
 NO & A & historical & -0.010 \\ 
    &  & fixed & -0.010 \\ 
    & B & historical & 0.038 \\ 
    &  & fixed & 0.038 \\ 
    & C & historical & 0.084 \\ 
    &  & fixed & 0.084 \\ 
    & D & historical & 0.100 \\ 
    &  & fixed & 0.100 \\ 
    & E & historical & 0.100 \\ 
    &  & fixed & 0.100 \\ 
   \hline
\end{tabular}
\caption{The table shows the value of the era-adjusted AB per HR of Barry Bonds in the 2001 season minus the era-adjusted AB per HR of Babe Ruth in the 1920 season under different configurations. The Park factor column indicates whether we apply the park-factor adjustment to the home run. The Talent pool column indicates the population changes we apply to the talent pool. Population A - E are constructed according to the five different estimates of the talent pool in Section 5 of the manuscript. The League size column indicates how full-time players are calculated. Historical means that we use the number of full-time players observed for the given. Fixed means that the number of full-time players is set to the maximum number of seasonal full-time players observed in our data set. The Difference indicates the value of the AB per HR of Barry Bonds in the 2001 season minus the AB per HR of Babe Ruth in the 1920 season under different configurations.}
\label{Tab:multiverseHRseason}
\end{table}

We also show how the distribution of the batting statistics can affect the overall batting average results. Table \ref{Tab:multiversenormal} shows the distribution of the batting statistics affects the top 25 career batting averages and top 25 four-year peaks by batting average. 

Stephen Jay Gould\footnote{\url{https://en.wikipedia.org/wiki/Full_House:_The_Spread_of_Excellence_from_Plato_to_Darwin}} suggests that the BA in every season follows a normal distribution \citep{Gould96} and we perform the Shapiro-Wilk Normality Test to verify this argument. The details are under Section \ref{batting} and we found the BAs do not follow the normal distribution for a significant proportion of seasons. So we use the nonparametric method to measure the BAs. We understand if the reader disagrees.

\begin{table}[ht]
\centering
\scalebox{0.65}{
\begin{tabular}{rlrlrlcrlcr}
  \hline
  & \multicolumn{4}{c}{Top 25 career BA} & \multicolumn{6}{c}{Top 25 four-year peak by BA}  \\
  & \multicolumn{2}{c}{nonparametric} & \multicolumn{2}{c}{parametric} & \multicolumn{3}{c}{nonparametric} & \multicolumn{3}{c}{parametric} \\
 & name & BA & name & BA & name & years &  BA & name & years & BA \\ 
  \hline
1 & Tony Gwynn & 0.342 & Tony Gwynn & 0.338 & Jose Altuve & 2014-2017 & 0.367 & Tony Gwynn & 1994-1997 & 0.360 \\ 
  2 & Rod Carew & 0.329 & Ty Cobb & 0.332 & Tony Gwynn & 1994-1997 & 0.366 & Ty Cobb & 1916-1919 & 0.353 \\ 
  3 & Jose Altuve & 0.327 & Rod Carew & 0.324 & Rod Carew & 1974-1977 & 0.363 & Wade Boggs & 1985-1988 & 0.351 \\ 
  4 & Ichiro Suzuki & 0.327 & Ichiro Suzuki & 0.322 & Miguel Cabrera & 2010-2013 & 0.355 & Rod Carew & 1974-1977 & 0.350 \\ 
  5 & Miguel Cabrera & 0.320 & Jose Altuve & 0.320 & Wade Boggs & 1985-1988 & 0.353 & Ichiro Suzuki & 2001-2004 & 0.348 \\ 
  6 & Roberto Clemente & 0.320 & Roberto Clemente & 0.318 & Ichiro Suzuki & 2001-2004 & 0.353 & Rogers Hornsby & 1921-1924 & 0.346 \\ 
  7 & Ty Cobb & 0.320 & Joe DiMaggio & 0.318 & Barry Bonds & 2001-2004 & 0.352 & Jose Altuve & 2014-2017 & 0.345 \\ 
  8 & Joe DiMaggio & 0.318 & Shoeless Joe Jackson & 0.316 & Joe Mauer & 2006-2009 & 0.350 & Mike Piazza & 1995-1998 & 0.345 \\ 
  9 & Wade Boggs & 0.316 & Wade Boggs & 0.314 & Roberto Clemente & 1964-1967 & 0.345 & Barry Bonds & 2001-2004 & 0.344 \\ 
  10 & Buster Posey & 0.316 & Freddie Freeman & 0.314 & Joe DiMaggio & 1938-1941 & 0.345 & Joe DiMaggio & 1939-1942 & 0.343 \\ 
  11 & Mike Trout & 0.315 & Stan Musial & 0.314 & Albert Pujols & 2003-2006 & 0.343 & Don Mattingly & 1984-1987 & 0.340 \\ 
  12 & Freddie Freeman & 0.314 & Ted Williams & 0.314 & Don Mattingly & 1984-1987 & 0.341 & Henry Aaron & 1956-1959 & 0.339 \\ 
  13 & Joe Mauer & 0.314 & Henry Aaron & 0.313 & Mike Piazza & 1995-1998 & 0.341 & Roberto Clemente & 1969-1972 & 0.338 \\ 
  14 & Ted Williams & 0.314 & Buster Posey & 0.313 & Willie Mays & 1957-1960 & 0.340 & Stan Musial & 1943-1946 & 0.338 \\ 
  15 & Stan Musial & 0.313 & Mike Trout & 0.312 & Matty Alou & 1966-1969 & 0.339 & Joe Mauer & 2006-2009 & 0.337 \\ 
  16 & Willie Mays & 0.312 & Matty Alou & 0.311 & Tim Anderson & 2019-2022 & 0.338 & Miguel Cabrera & 2010-2013 & 0.336 \\ 
  17 & Bill Terry & 0.312 & Miguel Cabrera & 0.311 & Stan Musial & 1943-1946 & 0.338 & Nap Lajoie & 1901-1904 & 0.336 \\ 
  18 & Robinson Cano & 0.311 & Robinson Cano & 0.311 & Rogers Hornsby & 1922-1925 & 0.335 & Albert Pujols & 2003-2006 & 0.336 \\ 
  19 & Henry Aaron & 0.310 & Vladimir Guerrero & 0.311 & Ted Williams & 1943-1946 & 0.335 & Matty Alou & 1966-1969 & 0.335 \\ 
  20 & Matty Alou & 0.310 & Joe Mauer & 0.311 & Ty Cobb & 1912-1915 & 0.334 & Honus Wagner & 1905-1908 & 0.335 \\ 
  21 & Vladimir Guerrero & 0.310 & Rogers Hornsby & 0.310 & Trea Turner & 2019-2022 & 0.334 & Tris Speaker & 1913-1916 & 0.334 \\ 
  22 & Derek Jeter & 0.310 & Willie Mays & 0.310 & Henry Aaron & 1956-1959 & 0.333 & Ted Williams & 1943-1946 & 0.334 \\ 
  23 & Al Oliver & 0.310 & Kirby Puckett & 0.310 & Cecil Cooper & 1980-1983 & 0.333 & Freddie Freeman & 2020-2023 & 0.332 \\ 
  24 & Lou Gehrig & 0.309 & Bill Terry & 0.310 & Freddie Freeman & 2020-2023 & 0.333 & Lou Gehrig & 1932-1935 & 0.332 \\ 
  25 & Edgar Martinez & 0.309 & Lou Gehrig & 0.309 & Nap Lajoie & 1901-1904 & 0.333 & Willie Mays & 1957-1960 & 0.332 \\ 
  \hline   
   \end{tabular}
   }
\caption{The table shows whether or not the distribution of the batting statistics is parametric or nonparametric affects the top 25 career batting averages and top 25 four-year peaks by batting average. The nonparametric indicates we use the nonparametric method to measure the batting averages. The parametric indicates we use the parametric method to measure the batting averages. }
\label{Tab:multiversenormal}
\end{table}

\section{Theoretical Results}
\label{empirical}






In the main text it was noted that $\widetilde{F}_{Y_i}(t)$ was explicitly constructed to be close to $\widehat{F}_{Y_i}(t)$. We now formalize this statement.

\begin{proposition}
	Let $\widetilde{F}_{Y_i}(t)$ be defined as in
(2) and let $\widehat{F}_{Y_i}(t)$ be the empirical distribution function. Then,
$
\sup _{t \in \mathbb{R}}\left|\tilde{F}_{Y_i}(t)-\widehat{F}_{Y_i}(t)\right| \leq \frac{1}{n}.
$
\end{proposition}

\begin{proof}
	We will prove this result in cases. First, when $t \leq \widetilde{Y}_{i,(1)}$ or $t \geq \widetilde{Y}_{i,(n+1)}$ we have that $\mid \widetilde{F}_{Y}(t)-$ $\widehat{F}_{Y}(t) \mid=0 .$ For any $j=1, \ldots, n$ and $\tilde{Y}_{i,(j)} \leq t<Y_{i,(j)},$ we have
$$
\left|\widehat{F}_{Y}(t)-\widetilde{F}_{Y}(t)\right|=\left|\frac{j-1}{n}-\frac{j-1+\left(t-\widetilde{Y}_{i,(j)}\right) /\left(\tilde{Y}_{i,(j+1)}-\tilde{Y}_{i,(j)}\right)}{n}\right| \leq \frac{1}{n}
$$
For any $j=1, \ldots, n$ and $Y_{i,(j)}<t<\tilde{Y}_{i,(j+1)},$ we have
$$
\left|\widehat{F}_{Y}(t)-\widetilde{F}_{Y}(t)\right|=\left|\frac{j}{n}-\frac{j-1+\left(t-\tilde{Y}_{i,(j)}\right) /\left(\tilde{Y}_{i,(j+1)}-\tilde{Y}_{i,(j)}\right)}{n}\right| \leq \frac{1}{n}
$$
Our conclusion follows.
\end{proof}

This leads to a Glivenko-Cantelli result which is appropriate for $\widetilde{F}_{Y}$.

\begin{corollary}
	Let $\widetilde{F}_{Y_i}(t)$ be defined as in (1) and let $\widehat{F}_{Y_i}(t)$ be the empirical distribution function. Then,
$
\sup _{t \in \mathbb{R}}\left|\widetilde{F}_{Y_i}(t)-F_{Y_i}(t)\right| \stackrel{a . s}{\longrightarrow} 0.
$
\end{corollary}

\begin{proof}
	We have, $\sup _{t \in \mathbb{R}}\left|\widetilde{F}_{Y}(t)-F_{Y}(t)\right| \leq \sup _{t \in \mathbb{R}}\left|\widetilde{F}_{Y}(t)-\widehat{F}_{Y}(t)\right|+\sup _{t \in \mathbb{R}}\left|\widehat{F}_{Y}(t)-F_{Y}(t)\right| .$ 
	The conclusion follows from the Glivenko-Cantelli Theorem and Proposition 2.1.
\end{proof}

\doublespacing
\bibliography{Gould96.bib}
\bibliographystyle{plain.bst}

\end{document}